\documentclass[11pt,a4paper]{article}
\usepackage{amsfonts,amssymb}

\newtheorem{define}{Definition}[section]
\newtheorem{lemma}[define]{Lemma}
\newtheorem{lemdef}[define]{Lemma and Definition}

\newtheorem{theorem}[define]{Theorem}
\newtheorem{corr}[define]{Corollary}
\newtheorem{example}[define]{Example}
\newtheorem{remark}[define]{Remark}
\newenvironment{indentedtext}[1]{\begin{description}
	\item[#1:]}{\end{description}}
\newenvironment{proof}{\begin{indentedtext}{Proof}}{\end{indentedtext}}

\newcommand{\NN}{\ensuremath{\mathbb{N}}}
\newcommand{\ZZ}{\ensuremath{\mathbb{Z}}}

\newcommand{\CC}{\ensuremath{\mathbb{C}}}
\newcommand{\PP}{\ensuremath{\mathbb{P}}}
\newcommand{\QQ}{\ensuremath{\mathbb{Q}}}
\newcommand{\OOne}{\mbox{\textup{1\hspace{-0.25em}l}}}
\newcommand{\LL}{\ensuremath{\mathcal{L}}}
\newcommand{\UU}{\ensuremath{\mathcal{U}}}
\newcommand{\WW}{\ensuremath{\mathcal{W}}}
\newcommand{\AAA}{\ensuremath{\mathcal{A}}}
\newcommand{\BBB}{\ensuremath{\mathcal{B}}}
\newcommand{\Nop}{\ensuremath{\mathcal{N}}}
\newcommand{\UW}{\UU_{\WW}}
\newcommand{\WWW}{\ensuremath{\WW(2,3^3)}}

\newcommand{\eps}{\varepsilon}
\newcommand{\xedarrow}{\nwarrow\hspace{-1em}\swarrow}

\newcommand{\p}{\ensuremath{\mathrm{p}}}
\newcommand{\spann}{\ensuremath{\mathrm{span}}}
\newcommand{\eigspace}{\ensuremath{\mathrm{eigenspace}}}

\newcommand{\Hom}{\ensuremath{\mathrm{Hom}}}
\newcommand{\skpi}[2]{\mbox{$\langle #1,#2\rangle$}}

\newcommand{\binom}[2]{\ensuremath{\left(\begin{array}{c} {#1} \\ {#2}%
    \end{array}\right)}}

\newcommand{\cmodl}{\ensuremath{\mathrm{Mod}_{\LL}}}
\newcommand{\cvec}{\ensuremath{\mathrm{Vec}_{\CC}}}
\newcommand{\cmodw}{\ensuremath{\mathrm{Mod}_{\WW}}}
\newcommand{\Ext}{\ensuremath{\mathrm{Ext}^1}}
\newcommand{\tr}{\ensuremath{\mathrm{tr}}}
\newcommand{\length}{\mbox{\textup{length}}}
\newcommand{\llength}{\mbox{\textup{LW-length}}}
\newcommand{\nlength}{\mbox{\textup{N-length}}}

\renewcommand{\[}{\begin{equation}}
\renewcommand{\]}{\end{equation}}
\newcommand{\refeq}[1]{equation (\ref{#1})}
\newcommand{\qed}{\hfill\checkmark}

\newcounter{saveequation}

\newenvironment{myeqnarray}%
{\stepcounter{equation}\setcounter{saveequation}{\value{equation}}%
\setcounter{equation}{0}%
\begin{eqnarray}}%
{\end{eqnarray}%
\setcounter{equation}{\value{saveequation}}%
\par\noindent\vskip-0.5ex}

\newenvironment{myeqnarraylabel}[1]%
{\refstepcounter{equation}%
\label{#1}%
\setcounter{saveequation}{\value{equation}}%
\setcounter{equation}{0}%
\begin{eqnarray}}%
{\end{eqnarray}%
\setcounter{equation}{\value{saveequation}}%
\par\noindent\vskip-0.5ex}

\begin{document}
\rightline{\vbox{\hbox{hep-th/9611160}\hbox{BONN-TH-96-17}
\hbox{November 1996}}}
\bigskip
\centerline{\LARGE On Reducible but Indecomposable Representations}
\smallskip
\centerline{\LARGE of the Virasoro Algebra}
\medskip
\centerline{Falk Rohsiepe\footnote{email: 
rohsiepe@avzw02.physik.uni-bonn.de}}
\centerline{Physikalisches Institut}
\centerline{der Universit\"at Bonn}
\centerline{Nu{\ss}allee 12, D-53115 Bonn}
\centerline{Germany}

\begin{abstract}
Motivated by the necessity to include so-called logarithmic operators in 
conformal field theories (Gurarie, 1993) at values of the central charge 
belonging to the logarithmic series $c_{1,p} = 1-6 (p-1)^2 / p$, reducible but 
indecomposable representations of the Virasoro algebra are investigated, 
where $L_0$ possesses a nontrivial Jordan decomposition.
After studying `Jordan lowest weight modules', where $L_0$ acts as 
a Jordan block on the lowest weight space (we focus on the rank two case),
we turn to the more general case of extensions of a lowest weight module by
another one, where again $L_0$ cannot be diagonalized. The moduli space of 
such `staggered' modules is determined. Using the structure of the moduli
space, very restrictive conditions on submodules of `Jordan Verma modules' 
(the generalization of the usual Verma modules) are derived. 
Furthermore, for any given lowest weight of a Jordan Verma module its
`maximal preserving submodule' (the maximal submodule, such that the quotient
module still is a Jordan lowest weight module) is determined.
Finally, the representations of the $\mathcal{W}$-algebra $\mathcal{W}(2,3^3)$
at central charge $c = -2$ are investigated yielding a rational logarithmic
model.
\end{abstract}

\section{Introduction}
Since the early works on 2-dimensional conformal field theory \cite{bpz}, the
representation theory of the Virasoro algebra $\LL$
\begin{myeqnarray}
\left[L_m, L_n\right] & = & (n-m) L_{m+n} +
        \frac{C}{12} \left(n^3-n\right) \delta_{n+m,0} \qquad 
        \forall n,m \in \ZZ \\
\left[ C, L_n \right] & = & 0 
        \qquad\qquad\qquad\qquad\qquad\qquad\qquad\qquad\qquad
        \forall n \in \ZZ
\end{myeqnarray}
\noindent\noindent
has been largely investigated using standard Lie algebra methods such as
lowest weight representations and irreducibility. The embedding
structure of lowest weight representations was resolved 
\cite{fefu1,fefu2,fefu4,fel,ffk} by close
examination of the Kac determinant \cite{kacdet}.
\par\noindent
Only recently it has been shown that for some values of the central charge
(when there are fields with integer spaced dimensions) the existence of fields
with logarithmic divergences in their four-point-functions is unavoidable
\cite{logarithms}. In fact this is true for the whole series of theories
on the edge of the conformal grid, namely if
$c = c_{1,q} = 1 - 6 (q - 1)^2 / q$,
$q \in \NN^{\ge2}$. Other CFTs exhibiting this logarithmic behaviour
are the WZNW model on the supergroup $GL(1,1)$ \cite{wznw}, gravitationally
dressed conformal field theories \cite{gravdress} and some critical
disordered models \cite{critdis1,critdis2}.
\par\noindent
These theories have physical relevance as they are supposed to describe aspects
of physical systems such as the fractional quantum Hall effect 
\cite{flohrhall,critdis2,hall2,hall},
2-dimensional polymer systems and random walks 
\cite{cardysaw,polymer1,polymer2}
or 2-dimensional turbulence \cite{flohrturb}.
In addition, $c = -2$ also appears in the theory of unifying
$\mathcal{W}$-algebras \cite{unify1,unify2,unify3}. Logarithmic conformal
field theories might also prove important for the description of
normalizable zero modes for string backgrounds 
\cite{string1, string2,string3}.
\par\noindent
Apart from the logarithmic behaviour of four-point-functions these theories
also exhibit a peculiar behaviour concerning their fusion structure:
If one defines the action of the Virasoro algebra on the tensor product of
two Virasoro representations in an appropriate way (see e.g.\ 
\cite{nahmfusion, gaberdiel, diplom}, also c.f.\ \cite{moorefusion}), 
starting with the set of ordinary lowest weight representations, 
one is naturally forced to include 
representations, where $L_0$ acts as a nontrivial Jordan cell on the lowest 
weight space.
In fact, representations of this kind were already found in
\cite{logarithms}. Therefore we will, after some general considerations,
focus on such representations, which are generated by two
vectors, on which $L_0$ acts as a nontrivial Jordan block.
We will call these representations \emph{Jordan lowest weight modules} 
(instead of the language of representations of an algebra from now on we 
will use the equivalent language of modules over an algebra). 
\par\noindent
Many of the results in this paper have already appeared in
\cite{diplom}; for a broader background the reader may refer to
this reference.
The paper is organized as follows:
After reviewing the most important facts from the theory of lowest
weight representations in section 2, the basic
definitions for our treatment of nondiagonal representations are given
in section 3.
In section 4 we proceed by studying the simplest examples of
representations of this kind, the above-mentioned
Jordan lowest weight modules. The submodules of 
Jordan lowest weight modules turn out to belong to an even broader
class of modules (we will call them \emph{staggered modules}), which
we will study in section 5. Sections 6 and 7 then turn back to
Jordan lowest weight modules and the classification of their submodules.
In section 8, we will generalize our definitions to \WW-algebras and study
the example of $\WWW$ at $c = -2$, which will turn out to be rational
in a slightly broadened sense. 
In section 9 we summarize the achieved results and point out 
directions for future research.

\section{Lowest weight modules revisited} \label{lowestsect}
The simplest class of modules of the Virasoro algebra $\LL$
is the class of lowest weight modules (LWMs). Though the structure
of these modules is well known since many years,
we will review the basic facts about them in this section.
There are two reasons for this. Firstly, we will present the facts
using a notation most suitable for our needs. Secondly,
our treatment of more complicated modules will sometimes 
be analogous to the lowest weight case, which we hope to
clarify by first presenting the known facts we will use
subsequently. The reader familiar with the theory of LWMs may skip this
section and directly turn to section \ref{generalcase} 
on page \pageref{generalcase}.
\par\noindent
Let $\UU$ denote the universal enveloping
algebra of $\LL$. As usual, let $\UU_k$ (``$k$-th level of $\UU$'') 
denote the span of monomials
of homogeneous degree $k$: $\langle L_{i_1}^{n_1} \ldots
L_{i_p}^{n_p} C^{n_c} ; n_i \in \NN^0, \sum_m n_m i_m = k \rangle$.
Furthermore $\UU^\pm \subset \UU$ and $\UU^0 \subset \UU$ will denote 
the universal enveloping algebras
of the subalgebras $\LL^\pm := \langle L_k; k \gtrless 0 \rangle$ 
and $\LL^0 := \langle L_0,C \rangle$ of the Virasoro algebra.
Hence, $\UU = \UU^- \UU^0 \UU^+$.

\begin{define}
A module $V$ of the Virasoro algebra is called
\textbf{lowest weight module (LWM)} if it contains a
subspace $W \subset V$ such that $\dim W = 1$ and
$V = \UU^+.W$.
\end{define}

\begin{define} \label{singulardef}
Let $V$ be an \LL-module. A vector $v \in V$ is called
\textbf{singular} if
\begin{enumerate}
\item $\forall n \in \NN: L_{-n} v = 0$
\item $L_0 v = h v;\; h \in \CC$
\item $C v = c v;\; c \in \CC$
\end{enumerate}
\end{define}

\begin{corr}
An \LL-module $V$ is a lowest weight module if and only if it contains
a singular vector $v \in V$ such that $V = \UU.v$.
The number $h$ in definition \ref{singulardef}
is then called the \textbf{lowest weight} and $v$ a
\textbf{lowest weight vector} of the module. 
\end{corr}

\begin{remark}
The central element $C\in\LL$ is, in general, represented by a linear operator.
But as it belongs to the center of $\LL$, in any irreducible 
representation this operator must be given
by a multiple of the identity operator $C = c \OOne$ (Schur's lemma).
The number $c$ is then called \textbf{central charge}.
In indecomposable representations this is also true as long as
$C$ is diagonalizable. In this paper we will not deal with other cases.
Therefore, we
always think of $C$ as a number. This even becomes necessary, if one
considers certain extensions of the Virasoro algebra, the so-called
\WW-algebras, which in general can only be consistently defined for
certain values of the central charge.
\par\noindent
As a consequence, we will sometimes be sloppy about the operator $C$
and treat it as a number right from the beginning. The scrupulous
reader may then e.g. substitute $\UU / \langle C-c\OOne \rangle$
for $\UU$.
\end{remark}
\begin{define}
A LWM $V$ with lowest weight $h$ and LWV $v$ is called 
\textbf{Verma module}, if it has the following universal 
property:
For any LWM $W$ with lowest weight $h$ and LWV $w$, there is
a unique $\LL$-homomorphism $V \rightarrow W$ mapping
$v$ to $w$.
\end{define}
\begin{theorem} \label{vermapbwbase}  
For any given $c, h \in \CC$, the Verma module $V(h,c)$
exists and is unique up to \LL-isomorphism. A base of the
module is given by
$$\left\{L_{i_1}\ldots L_{i_k} v | k \in \NN^0; 
i_1 \ge \ldots \ge i_k \right\},$$
where $v$ is the lowest weight vector.
\end{theorem}
\begin{proof}
Uniqueness is clear due to the universal property. The
existence is proven by construction:
$\UU$ itself is an \LL-module by left multiplication. Let
$V := \UU / \langle L_{-n} , (L_0-h\OOne), (C-c\OOne) \rangle$.
This obviously is a lowest weight module with lowest weight $h$ and
LWV $[\OOne]$. It is a Verma module by the universal property of $\UU$.
The last assertion follows from the Poincar\'e-Birkhoff-Witt theorem
for $\UU$.\hfill\checkmark
\end{proof}
\noindent Due to the universal property any lowest weight module is
(up to \LL-isomorphism) a quotient of a Verma module
by a proper submodule. We immediately deduce the following
\begin{corr}\label{uniqueirrlwm}
For any $h,c \in \CC$, there is an (up to isomorphism) uniquely determined
\textbf{irreducible} or \textbf{minimal} lowest weight module $M(h,c)$.
It is given by the quotient of $V(h,c)$ by its maximal proper
submodule.
\end{corr} 
\noindent It is a well known fact \cite{kacdet, fefu2}, that any submodule 
of a Verma module is generated by singular vectors and therefore is the sum 
of lowest weight modules. This immediately leads
to the question, which Verma modules can be embedded into a
given Verma module. This question may be answered using the
so-called Shapovalov form which we will define below:
\par\noindent
Given a Verma module $V(h,c)$ with LWV $v$ one can define a 
representation of $\LL$ (and thereby of $\UU$) on
its graded dual $V(h,c)^*$ by setting
\[(L_{i_1}^{n_1} \ldots L_{i_p}^{n_p} C^{n_c})^\dagger := 
C^{n_c} L_{-i_p}^{n_1} \ldots L_{-i_1}^{n_p}\label{invol}\]
and
\[((L_{i_1}^{n_1} \ldots L_{i_p}^{n_p} C^{n_c}) \phi)(w)
:= \phi((L_{i_1}^{n_1} \ldots L_{i_p}^{n_p} C^{n_c})^\dagger w)\]
where $\phi \in V(h,c)^*$ and $w \in V(h,c)$.
Let $V^\dagger(h,c) := \UU.v^* \subset V(h,c)^*$ denote the
\textbf{dual module} of $V(h,c)$. It obviously is a lowest
weight module with lowest weight $h$ and LWV $v^*$. The
\LL-homomorphism 
$$\left\{\begin{array}{lcl}
V(h,c) & \rightarrow  & V^\dagger(h,c) \\
u.v    & \mapsto      & u.v^*
\end{array}\right.$$
together with the natural pairing of $V(h,c)$
with $V(h,c)^*$ then yields a bilinear form $\skpi{.}{.}$ on
$V(h,c)$, the \textbf{Shapovalov form}.
As one easily checks by direct computation, the Shapovalov form is
symmetric and obeys 
\[k \not= l \Rightarrow \skpi{V(h,c)_k}{V(h,c)_l} = 0 \label{nointersect}\]
where $V(h,c)_k := \eigspace(L_0, h+k)$ is the $k$-th \textbf{level} 
of $V(h,c)$.
\par\noindent
One easily sees that the radical of $\skpi{.}{.}$ is exactly given by the
maximal proper submodule of $V(h,c)$, and therefore $V^\dagger(h,c) = M(h,c)$.
This fact also allows one to define the Shapovalov form on any LWM.
\par\noindent
Because of \refeq{nointersect} it makes sense to examine the determinant
of the restriction $\skpi{.}{.}_k$ of the Shapovalov form to the $k$-th level
of a given Verma module. A nontrivial intersection of the $k$-th level with the
maximal proper submodule may then be detected by the vanishing of the 
corresponding determinant. V.~Kac \cite{kacdet} gave an explicit formula for
this determinant, which was proven by B.L.~Feigin and D.B.~Fuks \cite{fefu1}:
\begin{theorem} The determinant $\det_n(h,c)$ of the matrix of $\skpi{.}{.}_n$
on $V(h,c)_n$ is given by
\[\label{kacdetformula}\begin{array}{l}
det_n(h,c) = K_n \prod\limits_{\stackrel{r,s \in \NN}{r s \le n}}
(h - h_{c;r,s})^{\p(n - r s)}, \\
h_{c;r,s} = \frac{1}{48} \left( (13-c)(r^2+s^2) +
\sqrt{(c-1)(c-25)}(r^2-s^2) - 24 r s - 2 + 2 c \right), \end{array}\]
where $\p(n)$ denotes the number of \textbf{partitions} of $n$ with generating
function
\[\left(\prod_{n \in \NN} (1-q^n)\right)^{-1} = \sum_{n \in \NN^0} \p(n) q^n,\]
and $K_n$ are nonvanishing constants (depending on the choice of base).
\end{theorem}
\noindent By careful examination of this formula B.L.~Feigin and D.B.~Fuks
were able to determine any Verma module that can be
embedded in a given one \cite{fefu2,fefu4}. To this end one parametrizes the
central charge by
\[c = 1 - 24 k,\]
which leads to
\[h_{r,s} = -k + {\textstyle \frac{1}{4}}\left((2k+1)(r^2+s^2) + 
2 \sqrt{k(k+1)}(r^2-s^2) - 2rs\right)\]
for the weights. Evidently, if $\nexists r,s: h = h_{r,s}$, the Verma module
$V(h,c)$ itself is irreducible. Using the convention $V_{r,s}:=V(h_{r,s},c)$,
the other, so-called \textbf{degenerate} cases can be classified as follows: 
\begin{theorem} \label{embedding}
Every degenerate representations of $\LL$ belongs to one of the following 
classes as determined by $k$, $k' := \sqrt{k(k+1)}$:
\begin{enumerate}
\item $k, k' \in \QQ$. In this case $k$ must be of the form 
      $\frac{(p-q)^2}{4pq}$ with $p,q \in \NN$ coprime,
      and therefore $c = 1 - 6\frac{(p-q)^2}{pq}$. In addition, one has
      $h_{r,s} \in \QQ\;\forall r,s \in \ZZ$. One distinguishes between three
      subcases:
      \begin{itemize}
      \item $q > p > 1$ (minimal models). We have $h_{r,s} = 
            \frac{(pr-qs)^2-(p-q)^2}{4pq}$.
            Based on the Verma modules $V_{r,s}$ with $1 \le r \le q-1$, $1 
            \le s \le p-1$, one has
            the following embedding lattices:
            \[\begin{array}{cccccccccc}
                    & \swarrow & V_{-r,s}   & \leftarrow & V_{r+2q,s} & 
                        \leftarrow & V_{-r-2q,s} & \leftarrow & V_{r+4q,s} & 
                        \cdots \\
            V_{r,s} & & & \xedarrow  & & \xedarrow & & \xedarrow & & \\
                    & \nwarrow & V_{2q-r,s} & \leftarrow & V_{r-2q,s} & 
                        \leftarrow & V_{4q-r,s}  & \leftarrow & V_{r-4q,s} & 
                        \cdots
            \end{array}\]
      \item $q > p = 1$ (logarithmic models). Here one has
            $h_{r,s}=\frac{(r-qs)^2-(q-1)^2}{4q}$. As is readily seen this set
            is already exhausted by the weights of the form $h_{r,1}$. Based 
            on the Verma modules $V_{r,1}$ with $r \in \{1,\ldots,q-1,q,2q\}$,
            we find the following embedding chains:
            \[{\setlength{\arraycolsep}{0.8\arraycolsep}\label{embedeq}
            \begin{array}{cccccccccccll}
            V_{r,1}  & =          & V_{2q-r,1} & \leftarrow & V_{2q+r,1} & 
                \leftarrow & V_{4q-r,1} & \leftarrow & V_{4q+r,1} & \cdots & 
                (r \notin q\NN) \\
            & & & & & & & & & & \\
            V_{r,1}  & \leftarrow & V_{r+2q,1} & \leftarrow & V_{r+4q,1} & 
                \leftarrow & V_{r+6q,1} & \leftarrow & V_{r+8q,1} & \cdots & 
                (r = q,2q) \\
            \end{array}}\]
      \item $p = q$, i.e. $c = 1$ (Gaussian models). The embedding structure
            for all degenerate modules is given by $V_{r,s} 
            \leftarrow V_{-r,s}$.
      \end{itemize}
\item $k \in \QQ$, $k' \in \CC\backslash\QQ$ (parabolic models, 
      c.f.~\cite{flohr4}). $c$ is still
      rational; the weights $h_{r,\pm r}\in\QQ\;\forall r \in\ZZ$ are exactly
      the rational weights.
      The embedding structure for all degenerate modules is
      $V_{r,s} \leftarrow V_{-r,s}$.
\item $k \in \CC\backslash\QQ$. Neither $c$ nor the weights (except for 
      $h_{1,1} = 0$) are
      rational. Again the embedding structure is $V_{r,s} \leftarrow V_{-r,s}$.
\end{enumerate}
\end{theorem}

\section{The general case} \label{generalcase}
For many physical applications, the knowledge of lowest weight modules is
completely sufficient. For example, in particle physics all
relevant representations must be unitary due to the
conservation of probability. Hence, all representations are
completely reducible and therefore a direct sum of
irreducible representations. With the additional constraint
of an energy spectrum bounded from below, irreducible
representations are automatically lowest weight (see
lemma \ref{irrlow} below) and the results of the
preceding section are completely satisfactory.
\par\noindent
Even in many statistical conformal field theories, where
unitarity plays a rather secondary role, one only has to deal
with lowest weight representations. 
\par\noindent
As mentioned before, only recently some cases
drew attention, in which this is not true anymore.
Even worse, in these cases $L_0$ is not diagonalizable,
but represented by matrices with a nontrivial
Jordan decomposition. We therefore must considerably
broaden the class of representations we want to deal
with. For thermodynamics to make sense we still put some 
restrictions on the class of representations we want to 
consider.
\par\noindent
In particular, the spectrum of $L_0$ must be discrete and
the real parts must be bounded from below 
($\tr e^{-\beta L_0}$ must exist). In fact, in all (mathematically)
interesting cases the spectrum will be real.
\par\noindent
As a consequence, 
$L_0$ possesses a Jordan decomposition $L_0 = L_0^d + L_0^n$
with $[L_0^d,L_0^n] = 0$, where $L_0^d$ is diagonalizable and 
$L_0^n$ operates nilpotently on its finite dimensional eigenspaces.
\par\noindent
For technical reasons we additionally demand $C$ to be diagonalizable.
\par\noindent
We will denote the category of Virasoro modules, which meet the
above restrictions, by $\cmodl$. Being a subcategory of the category 
$\cvec$ of complex vector spaces, it clearly is abelian. Its objects
will simply be called $\LL$-modules.
\par\noindent
Though this category is rather large compared to the category of
lowest weight modules, the situation is not as bad as one
would expect at first sight. Many of the properties of the
reducible but indecomposable representations we now have to
deal with can be played back to the properties of lowest weight
modules. The rest of this section is devoted to this aim.
\subsection{Gradation and filtration by $\mathbf{L_0}$}\label{gradfilt}
One easily computes
\[(L_0 - h - k)^m L_k = L_k (L_0 - h)^m,\label{prediagkomm} \]
and therefore
\[[L_0^d, L_k] = k L_k, \; [L_0^n, L_k] = 0\label{diagkomm}. \]
\noindent Now let $V \in \cmodl$. Clearly one has
\[V = \bigoplus\limits_{h \ge h_{min}} \eigspace(L_0^d, h).\]
Equation (\ref{diagkomm}) then implies
\[L_k: \eigspace(L_0^d, h) \rightarrow \eigspace(L_0^d, h+k).\]
For indecomposable $V$ one therefore has
\[V = \bigoplus\limits_{n=0}^{\infty} \eigspace(L_0^d, h_{min}+n).\]
\begin{define}
Let $V_n := \eigspace(L_0^d, h_{min} + n)$.
$V_n$ is called the \textbf{$\mathbf{n}$-th level} of $V$, $V_0$ is
also called its \textbf{lowest weight space}.
\end{define}
\begin{define}
$L_0^n$ induces a filtration
\[\bar{V}^{(1)} \hookrightarrow \bar{V}^{(2)} \hookrightarrow \ldots 
\hookrightarrow \bar{V}^{N} = V, \label{nilfilt}\]
where $\bar{V}^{(k)} := \ker((L_0^n)^k)$ and $N \in \NN \cup \{\infty\}$.
Because of \refeq{diagkomm} the $\bar{V}^{(k)}$ are submodules of $V$
and by definition $L_0$ operates diagonalizably on the factor modules
$V^{(k)} := \bar{V}^{(k+1)} / \bar{V}^{(k)}$.
$V^{(k)}$ is called the \textbf{$\mathbf{k}$-th stage} of $V$.
The number $N \in \NN \cup \{\infty\}$ is called the
\textbf{nilpotency length} of $V$ and is denoted by $\nlength(V)$.
\end{define}
\noindent For $N$ to be finite it is sufficient (but not necessary) that $V$ 
is finitely generated as a $\UU$-module.

\begin{define}\label{presdef} 
Let $V\in\cmodl$ be an \LL-module and $I\subset V$ a
submodule. $I$ is called \textbf{preserving} (the nilpotency length),
if $\nlength(V/I) = \nlength(V)$. $I$ is called
\textbf{maximal} preserving submodule, if there is no
preserving submodule $J \subset V$ with $I \subsetneq J$.
The module $V$ is called \textbf{minimal}, if it contains
no nonzero preserving submodules.
\end{define}

\subsection{More filtrations}
Apart from the filtration (\ref{nilfilt}) we want to introduce two more
filtrations. For every \LL-module $V \in \cmodl$ one has a chain of embeddings
\[V := V^0 \hookleftarrow V^1 \hookleftarrow V^2 \hookleftarrow \ldots,
\label{simplechain}\]
where $V^{k+1}$ is a maximal proper submodule of $V^k$.
The factor modules $M^k := V^k / V^{k+1}$ are irreducible.
\begin{define} \label{lengthdefine}
If the chain (\ref{simplechain}) ends, i.e. $\exists n \in \NN: V^n = 0$,
then the smallest $n\in\NN$ with this property is called the \textbf{length}
of the module and is denoted by $\length(V)$. Otherwise $\length(V) := \infty$.
\end{define}
\begin{lemma}\label{thereislwm}
For any \LL-module $V \in \cmodl$, there is a submodule $W \subset V$, which
is a lowest weight module.
\end{lemma}
\begin{proof}
Without loss of generality let $V$ be indecomposable. Let $V_0$ denote its
lowest weight space. There is at least one $v\in V_0$ which is
an eigenvector of $L_0$. Therefore $\UU.v \subset V$ is a lowest weight
module.\hfill\checkmark
\end{proof}
\begin{corr}
\label{irrlow} Any irreducible \LL-module $M\in\cmodl$ is a lowest 
weight module.
\end{corr}
\noindent As even the length of many Verma modules is $\infty$, we
need another, somewhat ``coarser'' measure of the complexity of
an \LL-module. To this end we use lemma \ref{thereislwm} and
examine sequences of the form
\[\label{lwmfilt}V =: W^0 \twoheadrightarrow W^1 \twoheadrightarrow W^2 
\twoheadrightarrow \ldots,\]
where $W^k = W^{k-1} / V^{k-1}$ and the $V^{k} \subset W^{k}$
are LWMs. Of course, in general there are arbitrarily many sequences of
this form, but nevertheless we can define:
\begin{define} \label{lwlength}
The smallest $n \in \NN$, for which there is a sequence of the form 
(\ref{lwmfilt}) with $W^n = 0$, is called \textbf{lowest weight length} of $V$
and is denoted by $\llength(V)$.
If there is no such $n \in \NN$ we set $\llength(V) := \infty$.
\end{define}

\begin{corr} The lowest weight length of an \LL-module $V$ is the
smallest integer $n$ such that $V$ contains a subspace $W \subset V$
with $\dim W = n$ and $V = \UU^+ W$.
\end{corr}

\begin{lemma} For any $V \in \cmodl$ one has
$$\nlength(V) \le \llength(V) \le \length(V).$$
\end{lemma}
\begin{proof}
If $\length(V) < \infty$, then any sequence
$$V = V^0 \hookleftarrow V^1 \hookleftarrow \ldots \hookleftarrow
V^n = 0$$
as in (\ref{simplechain}) induces a sequence
$$V = V / V^n \twoheadrightarrow V / V^{n-1} \twoheadrightarrow
\ldots \twoheadrightarrow V / V^1 \twoheadrightarrow V / V^0 = 0$$
and $V / V^k = (V / V^{k+1}) / (V^k / V^{k+1})$.
$V^k / V^{k+1}$ is irreducible and therefore according to corollary
\ref{irrlow} a LWM, which proves the second inequality.
Now consider a sequence
$$V = W^0 \twoheadrightarrow W^1 \twoheadrightarrow W^2 
\twoheadrightarrow \ldots$$
as in (\ref{lwmfilt}). As the nilpotency length of an LWM always is 1,
one either has $\nlength(W^k) = \nlength(W^{k-1})$ or
$\nlength(W^k) = \nlength(W^{k-1}) - 1$.\\ 
This proves the first inequality.\qed
\end{proof}

\noindent For future convenience we will now name the simplest cases:

\begin{define} \label{modulesdef}
An \LL-module $V\in\cmodl$ with
$\nlength(V) = \llength(V) = k, k\in\NN^{\ge2}$ is called
\textbf{staggered module}. The number $k$ is called its
\textbf{rank}.
\par\noindent
If a staggered module $V$ contains a subspace $W \subset V$ 
with $\dim W = 1$ and $V = \UU^0 \UU^+ W$, $V$ is called
\textbf{Jordan lowest weight module (JLWM)}.
\par\noindent
If a staggered module $V$ with rank $k$ contains a subspace
$W \subset V$ with $\dim W = k$ and $V = \UU^+ W$, such that
$W = \bigoplus_{n = 1}^k W_n$, $\forall n: \dim W_n = 1$, 
$L_0^d W_n = \lambda_n W_n$ and 
$n \not= m \Rightarrow \lambda_n \not= \lambda_m$,
$V$ is called \textbf{strictly staggered}.
\end{define}

\noindent For the rest of this paper we will restrict ourselves to the
so-called \emph{logarithmic models} with central charge
$c_{1,q}, q\in\NN^{\ge 2}$ (see theorem \ref{embedding}).
There are three reasons for this:
\begin{itemize}
\item Firstly, for these theories one has towers of weights with integer
spacing, so that following V.~Gurarie \cite{logarithms}
one has to introduce representations with nilpotency length $>1$
in order to guarantee the consistency of OPE and conformal blocks.
The necessity to do so can also be seen when calculating the fusion product
of two LWMs \cite{kauschfusion,diplom}.
\item Secondly, theories with these central charges have applications
in various fields of physical and mathematical interest as e.g.
the fractional quantum hall effect \cite{flohrhall,hall}, the two-dimensional 
polymer system and 2D random walks \cite{polymer1,polymer2,cardysaw}, 
turbulence \cite{flohrturb}
and the theory of unifying \WW-algebras \cite{unify1,unify2,unify3}.
\item The third reason is a rather technical one: The comparatively
simple embedding structure of Verma modules (\refeq{embedeq}) as compared
to the minimal models significantly simplifies the study of modules 
with nilpotency length $>1$.
\end{itemize}

\section{Jordan lowest weight modules}
We first investigate the simplest case of modules with
nilpotency length $>1$, namely the above defined
Jordan lowest weight modules (the first
example studied by V.~Gurarie in \cite{logarithms} was of this type).
Their treatment is largely simplified by the following
\begin{lemdef} 
An \LL-module $M$ is a JLWM of rank $k$, if and only if
there are $k$
linearly independent vectors $v^{(0)},\ldots,v^{(k-1)} \in M$,
such that the following conditions are fulfilled:
\begin{enumerate}
\item ${C.v = c\,v\qquad\forall v \in M}$
\item ${L_0.v^{(n)} = h\,v^{(n)} + v^{(n-1)}}\;\;\forall n \in
        \{1,\ldots,k-1\}    \quad\mbox{and}\quad
        L_0.v^{(0)} = h\,v^{(0)}$
\item ${L_{-n}.v^{(m)} = 0\qquad\forall n \in \NN, m \in \{0,\ldots,k-1\}}$
\item $M = \UU.v^{(k-1)}.$
\end{enumerate}
$h$ is called \textbf{lowest weight} of the module and the $v^{(n)}$ its
\textbf{lowest weight vectors}. If $k = 2$, $v^{(1)}$
is called \textbf{upper} and $v^{(0)}$ \textbf{lower} lowest weight vector.
\end{lemdef}
\begin{proof}
If the module $M$ fulfills the above conditions,
a subspace $W$ as in definition \ref{modulesdef} is
given by $W := \spann(v^{(k-1)})$.
Conversely, let $M$ be a JLWM of rank $k$ and let $W \subset M$ be a 
one-dimensional subspace
as in definition \ref{modulesdef}.
Then choose $0 \not= v^{(k-1)} \in W$. Further let
$\forall n \in \{k-2,\ldots,0\}: v^{(n)} := L_0^n v^{(n+1)}$. 
\qed
\end{proof}

\noindent For simplicity we will further restrict ourselves to the
rank 2 case. Nevertheless, most of the results are analogously valid
for higher ranks. The necessary modifications are almost always obvious.
\par\noindent
In analogy to the lowest weight case and define:
\begin{define} A Jordan lowest weight module $V$ with lowest weight $h$ and
lowest weight vectors $v^{(0)}, v^{(1)}$ is called
\textbf{Jordan Verma module (JVM)}, if it fulfills the following
universal property:
For any \textup{JLWM} $W$ with lowest weight $h$ and lowest weight vectors
$w^{(0)}, w^{(1)}$, there exists a unique \LL-homomorphism
$V \rightarrow W$ mapping $v^{(0)}$ to $w^{(0)}$ and $v^{(1)}$ to $w^{(1)}$ .
\end{define}
\begin{theorem}\label{existjvm}  
For any given $h,c \in \CC$, the Jordan Verma module $\tilde{V}(h,c)$
exists and is uniquely determined up to \LL-isomorphism.
\end{theorem}
\begin{proof}
As before, uniqueness is clear due to the universal property. Again the
existence is proven by construction:\par\noindent
$\hat{V} := \UU \times \UU$ is an \LL-module by left multiplication.
Let $M \subset \hat{V}$ denote the left ideal generated by
$\left\{(L_0-h,-1),(0,L_0-h),(0,L_{-n}),(L_{-n},0);n\in\NN\right\}$.
Evidently, $\tilde{V}(h,c):= \hat{V} / M$ is the wanted JVM (with
lowest weight vectors $[(\OOne,0)]$ and $[(0,\OOne)]$).\qed
\end{proof}
\begin{remark}
Alternatively one could have divided $\UU$ by the left ideal
$\left\langle (L_0-h)^2,L_{-n};n\in\NN\right\rangle$, thereby obtaining lowest
weight vectors $[L_0-h]$ and $[\OOne]$. Equivalence is easily proven using
the Poincar\'e-Birkhoff-Witt theorem.
\end{remark}
\begin{define}
Let $V$ be a JLWM of rank 2. Then $V^{(0)} := \ker L_0^n$ is called 
\textbf{lower} and $V^{(1)} := V / V^{(0)}$
\textbf{upper module} of $V$.
\end{define}
\begin{corr}\label{updownverma}
Let $V, V^{(0)}$ and $V^{(1)}$ be as above. In $\cvec$ one has 
$V = V^{(0)} \oplus V^{(1)}$. If $V$ is a JVM, then
$V^{(0)}$ and $V^{(1)}$ are Verma modules.
\end{corr}
\begin{proof} 
See the proof of theorem \ref{existjvm}.\qed
\end{proof}
\begin{corr}
Let $\tilde{V}(h,c)$ be the Jordan Verma module with lowest weight $h$
and lowest weight vectors $v^{(0)}$ and $v^{(1)}$. 
Then a base of $\tilde{V}(h,c)$ is given by
\[\left\{L_{k_n}\ldots L_{k_1}.v^{(j)}; n\in\NN^0, j\in\{0,1\},
0 < k_1 \le \ldots \le k_n\right\}.\]
\end{corr}
\begin{proof}
See corollary \ref{updownverma} and theorem \ref{vermapbwbase}.\qed
\end{proof}
\subsection{Shapovalov form and minimal JLWMs}
Let $V = \tilde{V}(h,c)$ be the JVM with lowest weight $h$, central charge $c$
and lowest weight vectors $v^{(0)},\ldots,v^{(k-1)}$.
As before the graded dual $V^{*}$ becomes an \LL-module by setting
\[\left.\begin{array}{ll}
(L_n \phi)(w) := & \phi (L_{-n}.w) \\
(C \phi)(w) := & \phi (C.w) \end{array}
\;\right\}\;\forall \phi \in V^{*}, v \in V.\]
Let $(v^{(n)})^{\dag}.v^{(m)} := \delta_{m,n}$,
$(v^{(n)})^{\dag}.w := 0 \;\forall w \in V_k, k > 0$. 
One calculates
\[(L_0.(v^{(n)})^{\dag}).v^{(m)} = (v^{(n)})^{\dag}.L_0.v^{(m)}
= h \delta_{n,m} + \delta_{n,m-1} = (h (v^{(n)})^{\dag} + 
(v^{(n+1)})^{\dag}).v^{(m)}.\]
$V^{\dag} := \UU.(v^{(0)})^{\dag}$
is a JLWM with lowest weight $h$, lowest weight vectors
\[\left\{v^*_{(n)} := (v^{(k-1-n)})^{\dag};\;n\in \{0,\ldots,k-1\}\right\}\]
and central charge $c$. Just as before, the \LL-homomorphism
\[\label{jordanfactormap} \phi: \left\{
\begin{array}{lll}
V & \rightarrow & V^{\dag} \\
u.v^{(k-1)} & \mapsto & u.v^*_{(k-1)} \;\;\forall u \in \UU^+
\end{array} \right.\]
together with the natural pairing between $V$ and $V^*$
induces the symmetric Shapovalov form $\skpi{.}{.}$ on $V$.
\par\noindent
If, on the other hand, one starts with the claim for symmetry
and contravariance with regard to the involution
$L_n \mapsto L_{-n}$, one is led to the same form
(there is some freedom of choice which stems from the 
selection of lowest weight vectors).
\par\noindent
The radical of $\skpi{.}{.}$ now obviously is \emph{not} the
maximal proper submodule of $V$ ($v^{(0)} \not\in Rad(V)$). 
In fact, the quotient of a JVM with lowest weight $h$ by its maximal proper
submodule is comparatively uninteresting, as by lemma \ref{irrlow} it is
just the irreducible LWM $M(h,c)$.
A more interesting analogue to maximal proper submodules and
irreducible factor modules is given by maximal preserving
submodules and minimal factor modules (definition \ref{presdef}).
\begin{lemma}
Let $V$ be a JLWM with lowest weight $h$. A submodule $I \subset V$ is 
preserving, if and only if $I \cap V_0 = 0$.
\end{lemma}
\begin{proof}
If $I \cap V_0 \not= 0$, it follows that $V^{(0)} \subset I$ and therefore 
$M/I$ has smaller nilpotency length than $V$. The other direction is clear.
\qed
\end{proof}
\begin{corr}
The maximal preserving submodule of a JVM is uniquely determined. It is
given by the union of all preserving submodules.
\end{corr}
\begin{corr}
The minimal JLWM for given lowest weight $h$ exists and is uniquely determined
up to \LL-isomorphism. It is given by 
$\tilde{M}(h,c) := \tilde{V}(h,c) / I_{max}$, where
$\tilde{V}(h,c)$ is the JVM with lowest weight $h$ and $I_{max}$ its maximal
preserving submodule.
\end{corr}
\noindent
One easily sees, that the radical of the Shapovalov form on a given JVM
is just its maximal preserving submodule.
Using the universal property it is clear, that any JLWM with lowest weight $h$
is isomorphic to a factor module $\tilde{V}(h,c) /I$, where $I$ is a
preserving submodule of $\tilde{V}(h,c)$.
Hence, the Shapovalov form is well defined for any JLWM and its radical
always is the maximal preserving submodule.
Furthermore, the Shapovalov form is nondegenerate on a minimal JLWM.
\par\noindent
Unfortunately the determinant of the Shapovalov form does not prove to be
as useful as in the lowest weight case: One easily calculates, that the matrix
$A_n$ of the restriction $\skpi{.}{.}_n$ of the Sha\-po\-va\-lov form
to the $n$-th level of $\tilde{V}(h,c)$ is given by
\[A_n = \left(\begin{array}{cc}
0 & \tilde{A}_n \\ \tilde{A}_n & *\end{array}\right),\]
where $\tilde{A}_n$ is the matrix of the restriction of the Shapovalov form
to the $n$-th level of the Verma module $V(h,c)$.
Its determinant therefore is minus the square of the determinant of 
$\tilde{A}_n$, and its zeroes consequently don't provide any new information
about the possible preserving submodules of $\tilde{V}(h,c)$. This was to be
expected, since any proper
submodule of $\tilde{V}(h,c)^{(0)} = V(h,c)$ is a preserving
submodule of $\tilde{V}(h,c)$.
In order to determine every possible preserving submodule of $\tilde{V}(h,c)$,
we therefore have to use other means than the Shapovalov form.
\subsection{Submodules of JVMs}
For the sake of simplicity we again restrict ourselves to the
case of nilpotency length 2 at central charge $c = c_{1,q}$.
Obviously the only interesting cases are the modules
$\tilde{V}(h,c)$, where $h = h_{r,s}$ as in theorem
\ref{embedding} with $r,s \in \NN$:
\begin{lemma} \label{noseriesnosub}
Let $\tilde{V}(h,c)$ be the JVM with lowest weight $h$ and 
$h \not= h_{r,s}\;\forall r,s\in\NN$
(see theorem \ref{embedding}). Then $\tilde{V}(h,c)$ contains no nonzero
proper submodules.
\end{lemma}
\begin{proof}
Suppose $J \subset \tilde{V}(h,c)$ is a proper submodule. Then
$J^{(0)} := J \cap \tilde{V}(h,c)^{(0)}$ and $J^{(1)} := J / J^{(0)}$ 
are submodules of $\tilde{V}(h,c)^{(0)}$ and $\tilde{V}(h,c)^{(1)}$,
respectively. 
As $h \not= h_{r,s}\;\forall r,s\in\NN$, $J^{(0)}$ and $J^{(1)}$ contain no
proper submodules and therefore $J = \tilde{V}(h,c)$, which contradicts the 
assumption of $J$ being a proper submodule.\qed
\end{proof}
\noindent In order to prevent unnecessary repetition, we will now fix some 
notations for the rest of this paper:
\begin{define}\label{notation}
Let $c = c_{1,q}, q \in \NN^{\ge 2}$, and let $V := V(h_{r,s},c)$ be the 
Verma module with lowest weight $h_{r,s}$, $r,s \in \NN$. Then we
denote by 
$$V(h_{r,s},c_{1,q}) =: V^1 \hookleftarrow V^2 \hookleftarrow V^3 
\hookleftarrow V^4 \ldots$$
the chain of embeddings according to theorem \ref{embedding}.
Furthermore, let $h^k$ be the lowest weight of $V^k$ and $V^{\infty} := 0$.
In addition, we choose lowest weight vectors $v^k \in V^k$. Then let
$\chi_k \in \UU^+$, such that $\chi_k.v^k = v^{k+1}$.
\end{define}
\noindent We now proceed by searching restrictions which must be met 
by submodules of a given JVM.
\begin{lemma} \label{firstclassify}
Let $\tilde{V} = \tilde{V}(h_{r,s},c)$ denote a JVM and other notations as 
defined above. 
Let $J \subset \tilde{V}$ be a submodule.
As in lemma \ref{noseriesnosub} $J^{(0)} := J \cap \tilde{V}^{(0)}$ and 
$J^{(1)} := J / J^{(0)}$ are submodules of $\tilde{V}^{(0)}$ and
$\tilde{V}^{(1)}$ respectively.
Then $J^{(1)} = V^n$ and $J^{(0)} = V^m$ with $n,m \in \NN \cup \{\infty\}$.
\end{lemma}
\begin{proof} 
Clear.\qed 
\end{proof}
\begin{lemma} \label{restrict}
Let $\tilde{V}$, $J$, $J^{(1)} = V^n$ and
$J^{(0)} = V^m$ as above. Then $n \ge m$.
If $J$ is a proper submodule, then $n > 1$. If $J$ is preserving, then $m > 1$.
\end{lemma}
\begin{proof}
Let $v^{(1)} + J^{(0)}$, $v^{(1)} \in J$, be a lowest weight vector of 
$J^{(1)}$. Trivially one has $k < l \; \Leftrightarrow \; h^k < h^l$.
Therefore $0\not=(L_0 - h^n).v^{(1)} \in \tilde{V}_{h^n - h^1} \cap J^{(0)} \;
\Rightarrow \; m \le n$. The rest is clear.\qed
\end{proof}
\noindent
Even more is true --- the submodule $J$ is already completely fixed by the two
integers $n$ and $m$, but due to a lack of notation we postpone this
result to the next section (lemma \ref{weightsareall}).
\par~\par\noindent
The restrictions imposed by lemma \ref{restrict} on a submodule of a given
JVM are necessary, but in general not sufficient for its existence.
This becomes clear as one studies the following examples:
\begin{example} \label{nojordan}
Let $c = c_{1,q} = 1 - 6{{{\left( q - 1 \right)}^2} \over q}$, $q \in
\{2,3,\ldots\}$, and $\tilde{V}$ be the JVM with lowest weight $0$ and lowest 
weight vectors $v^{(0)}, v^{(1)}$. Let $J$ denote its maximal preserving
submodule. 
Then $J^{(0)} = V^2$ is the Verma module with lowest weight 1. 
We will now show, that $J^{(1)} = V^n$ with $n > 2$. Assume, that 
$J^{(1)} = V^2$ with lowest weight vector $L_1.v^{(1)} + J^{(0)}$. 
One easily calculates
\begin{eqnarray*}
L_{-1}.(L_1.v^{(1)} + \alpha L_1.v^{(0)})& = &
[L_{-1},L_1].(v^{(1)} + \alpha v^{(0)}) \\ & = &
2 L_0.(v^{(1)} + \alpha v^{(0)}) \\ & = & 2 v^{(0)}.
\end{eqnarray*}
Therefore, $J^{(0)} = V^{(0)}$, which contradicts the assumption of $J$
being preserving.\qed
\end{example}
\begin{example} \label{firststaggered}
Now let $q = 2$ ($c = -2$). We will show that $J^{(1)} = V^3$: Let
$$w := (2 L_2 L_1 - L_1^3).v^{(1)} - \tilde{w},\;\; \tilde{w} \in V^{(0)}_3.$$
In $V / J^{(0)}$ one easily calculates ($[w] := w + J^{(0)}$ being 
the equivalence class of $w$):
$$L_0.[w] = 3 [w].$$
The system of equations
$$L_{-k}.[w] = 0, \qquad k \in \{1,2,3\}$$
is uniquely solved by
$$[\tilde{w}] = [L_3.v^{(0)}].$$
In addition,
$$L_{-2}.w = -15 L_1.v^{(0)}.$$
Therefore
$$J = \UU.w,\quad J^{(0)} = V^2\quad \mbox{and} \quad J^{(1)} = V^3.\qed$$
\end{example}
\noindent
The above examples suggest that for any JVM with lowest weight $h_{r,s}$ 
there always exists a preserving submodule $J$ with
$J^{(0)} = V^2$ and $J^{(1)} = V^3$. Furthermore one might assume that
there never is a submodule $J$ with $J^{(0)} = J^{(1)} = V^2$.
The first statement will prove to be true, while for the second one
we will find counterexamples.

\section{Staggered modules}
The submodule $J$ in example \ref{firststaggered} is neither a JLWM
nor a lowest weight module, nor is it the direct sum of such.
It belongs to the broader class of staggered \LL-modules defined in
definition \ref{modulesdef}, where the nontrivial Jordan
decomposition of $L_0$ shows but at higher levels.
In fact, M.R.~Gaberdiel and H.G.~Kausch \cite{kauschfusion} found
modules of this kind in the fusion product of lowest weight modules at $c=-2$,
which do \emph{not} occur as submodules of JLWMs.
We therefore leave the submodule point of view and extend our investigations 
to general modules of this form (we will again restrict ourselves to the 
rank 2 case). After all, this will also prove useful for the
classification of the maximal preserving submodules of JVMs.
\begin{lemdef} \label{staggereddef}
Let $c = c_{1,q} = 1 - 6{{{\left( q - 1 \right)}^2} \over q }$.
An \LL-module $M$ with nilpotency length $\nlength(M) = 2$ is a 
staggered module if and only if there is
a pair of vectors $v^{(1)}, v^{(0)} \in M$ and numbers 
$h^{(1)}, h^{(0)} \in \CC$, 
such that the following conditions are met:
\begin{enumerate}
\item $V^{(0)} := \UU.v^{(0)}$ is LWM with LWV $v^{(0)}$ and lowest 
weight $h^{(0)}$.
\item $V^{(1)} := M / V^{(0)}$ is a LWM with LWV $v^{(1)} + V^{(0)}$ 
and lowest weight $h^{(1)}$.
\item $0 \not= (L_0-h^{(1)}).v^{(1)} =: v_0 \in
      V^{(0)}_{h^{(1)}-h^{(0)}}$.
\item $L_{-1}.v^{(1)} =: v_1 \in
      V^{(0)}_{h^{(1)}-h^{(0)}-1}$.
\item $L_{-2}.v^{(1)} =: v_2 \in
      V^{(0)}_{h^{(1)}-h^{(0)}-2}$.
\end{enumerate}
$h^{(1)}$ ($h^{(0)}$) is called \textbf{upper (lower) lowest weight}
and $v^{(1)}$ ($v^{(0)}$) \textbf{upper (lower) lowest weight vector}.
The module is strictly staggered, if $h^{(1)} > h^{(0)}$.
\end{lemdef}
\begin{proof}
clear.\qed
\end{proof}

\begin{lemma} \label{mustprime}
Let $M$ be a staggered module, $h^{(0)}$, $h^{(1)}$, $v^{(0)}$,
$v^{(1)}$, $v_0$, $v_1$ and $v_2$ as defined above. Then $v_0$ is singular.
\end{lemma}
\begin{proof}
One calculates
\[\begin{array}{ll}
L_{-1}.v_0 & = L_{-1}(L_0-h^{(1)}).v^{(1)} = (L_0-h^{(1)}).v_1 + 
[L_{-1},L_0].v^{(1)} \\
& = (L_0-h^{(1)}).v_1 + v_1 = 0.\end{array}\]
Analogously one gets $L_{-2}.v_0 = 0$. With
\[\label{togethercomplete}\lbrack \underbrace{L_{-1} \lbrack L_{-1} \ldots
\lbrack L_{-1}}_{\mbox{$n$ times}}
,L_{-2} \rbrack ... \rbrack\rbrack = (-)^n n! L_{-2-n}\]
the assertion follows.\qed
\end{proof}
\noindent
We immediately get the
\begin{corr} \label{weightrestrict}
For $h^{(1)}, h^{(0)} \notin \{h_{r,s}; r,s \in \NN\}$, no proper staggered 
modules can exist. More precisely:
Let $V^k$ and $h^k$ be as in definition \ref{notation}. 
For given lower lowest weight $h^k$ only staggered modules with 
upper lowest weight $h^m$,$m \ge k$, can exist.
Of course, at least one such module always exists, namely the corresponding
submodule of the JVM $\tilde{V}(h^k, c)$.
\end{corr} 
\begin{corr}
Let $V$ be a staggered module with lowest weights $h^{(0)}$ and $h^{(1)}$.
A submodule $I \subset V$ is preserving, 
if and only if $I \cap V_{h^{(1)}-h^{(0)}} = 0$. Hence, the maximal preserving
submodule of $V$ is uniquely determined.
\end{corr}
\begin{define} A staggered module $V$ is called \textbf{vermalike}, if it
fulfills the following universal property:
If $M$ is a staggered module with the same lowest weights and
$\phi: M \rightarrow V$ is an \LL-epimorphism, then $\phi$ is an
\LL-isomorphism.
\end{define}
\begin{remark}
Compared to Verma modules and JVMs we defined the universal property
in this case ``the other way around'', as it is true that nonisomorphic 
vermalike modules always have different $\{v_0,v_1,v_2\}$, but the
converse is not true.
\end{remark}
\noindent
With the above notations, we now return to submodules of JVMs and
prove, that the numbers $n$ and $m$ of lemma \ref{restrict} already
fix the corresponding submodule completely.
\begin{lemma} \label{weightsareall}
Let $\tilde{V}$ denote the JVM with lowest weight $h$ with notations as
in definition \ref{notation}.
Let $I \subset \tilde{V}$ and $J \subset \tilde{V}$ be staggered submodules 
with $I^{(1)} \cong J^{(1)}$ and $I^{(0)} \cong J^{(0)}$. Then $I = J$.
\end{lemma}
\begin{proof}
Clearly $I^{(0)} = J^{(0)}$ (theorem \ref{embedding}).
We assume, that $I \not= J$.
Let $v_I$ and $v_J$ be upper LWVs of $I$ and $J$ respectively. Furthermore
choose $w = \alpha v_I - v_J$ so that $w+\tilde{V}^{(0)} = 0+\tilde{V}^{(0)}$
(this is always possible, as theorem \ref{embedding} forces the LWVs of
$I^{(1)}\cong J^{(1)}\subset M^{(1)}$ to be scalar multiples of one another). 
Because of $I \not= J$ we have $w \notin I^{(0)}$ (otherwise, $v_J \in I$).
As $v_I+I^{(0)}$ and $v_J+I^{(0)}$ are singular in $M/I^{(0)}$, this also 
applies to $w+I^{(0)}$. Now $\tilde{V}^{(0)}/I^{(0)}$ contains no nonzero 
singular vectors on the level in consideration and therefore the assumption
must be false.\qed
\end{proof}
\begin{remark}
An analogous statement obviously applies even when $\tilde{V}$ is no JVM but a
proper staggered module.
\end{remark}
\subsection{Moduli spaces} \label{modspaces}
By definition any staggered module is given as the quotient of a vermalike
staggered module by a preserving submodule (a preserving submodule,
as before, is a submodule such that the factor module has the same
nilpotency length). Therefore we are interested in how
many nonisomorphic staggered modules exist for given lowest weights.
To this end we will first discuss, which choices of $v_0, v_1$ and $v_2$
occur as data of staggered modules.
\par~\par\noindent
With the notations from definition \ref{notation} we want to study all possible
vermalike staggered modules with lowest weights $h^{(0)} = h^1$ and
$h^{(1)} = h^k$, $k > 1$.
Suppose, that there exists a vermalike staggered module $M$ with given data
$[v_0], [v_1], [v_2]$, where $v_0, v_1, v_2 \in V^1$ and
$[v]$ denotes the image of $v$ under the surjection
$V^1 \rightarrow M^{(0)}$. Then by the universal properties of $\UU$ and $V^1$
this vermalike staggered module can be constructed as follows:
\par\noindent
Let $V := \UU \oplus V^1$ where $\UU$ is an \LL-module by left multiplication.
Furthermore let $I \in V$ denote the left ideal generated by 
$\left\{ (L_{-1}, -v_1), (L_{-2}, -v_2),
(L_0 - h^{(1)}, -v_0) \right\}$, and let $M := V / I$.
Evidently, this must be the wanted vermalike staggered module.
\par\noindent
Now the question, whether the choices $[v_0], [v_1], [v_2]$ occur as data of
a vermalike staggered module, reduces to the question, whether the above
constructed module $M$ is a staggered module. In particular, we must have
$[v_0] \not= 0$ and we obtain the
\begin{lemma} \label{vermalikemeansverma}
Let $M$ be a vermalike staggered module with lowest weights $h^1$ and $h^k$.
Then $M^{(0)}$ and $M^{(1)}$ are Verma modules, i.e.
$M^{(0)} = V^1$ and $M^{(1)} = V^k$.
\end{lemma}
\begin{proof}
See the above construction.\qed
\end{proof}
\noindent
We now want to study, under which circumstances $M$ fails to be staggered.
We first concentrate on the case $k = 2$ and define $N_2 := h^2-h^1$. 
$M$ fails to be staggered if and only if $(0,v^1) \in I$. This is equivalent to
$$\exists \tilde{\xi}, \tilde{\psi} \in \UU:\;
(\tilde{\xi}.L_{-1} - \tilde{\psi}.L_{-2}, -\tilde{\xi}.v_1 +
\tilde{\psi}.v_2) = (0, v^1).$$
\noindent If one expands the left hand side of this equation in terms of a PBW
base of $\UU$, where negative modes are sorted to the right, this turns out to
be equivalent to
$$\exists \xi, \psi \in \UU^-:\;
(\xi.L_{-1} - \psi.L_{-2}, -\xi.v_1 + \psi.v_2) = (0, v^1).$$
\noindent Now let $\UU^-_n$ be the $n$-th level of $\UU^-$ (i.e. the linear 
span of monomials
$L_{-k_1} \ldots L_{-k_m}$ with $\sum_{i=1}^m k_i = n$).
Further let $S := \UU^-.L_{-1} \cap \UU^-.L_{-2} \subset \UU^-$. We now want to
determine $\dim S_n$ ($S_n := S \cap \UU^-_n$).
\par\noindent
We first remark, that $\UU^-.L_{-1} + \UU^-.L_{-2} = \UU^-$,
which follows from \refeq{togethercomplete}.
Hence, \[\dim S_n = \dim (\UU^-.L_{-1})_n + \dim (\UU^-.L_{-2})_n -
 \dim \UU^-_n = \p(n-1) + \p(n-2) - \p(n)\label{dimexample}.\]
\par\noindent
We define a linear map
\[\phi: \left\lbrace \begin{array}{rcl}
S & \longrightarrow & \UU^- \times \UU^-\\
\psi = \psi^1.L_{-1} = \psi^2.L_{-2} & \longmapsto & (\psi^1 , \psi^2)
\end{array} \right.\]
and then define $\tilde{L} := \phi(S)$ and $\tilde{L}_n := \phi(S_n)$.
$\phi$ is well defined (Poincar\'e-Birkhoff-Witt for $\UU^-$) and
a vector space isomorphism, whence $\dim \tilde{L}_n = \dim S_n$.
Now we define
$\rho: \tilde{L}_n \rightarrow \Hom(V^1_{m-1} \times V^1_{m-2}, V^1_{m-n})$
by setting
$$\rho\left((\psi^1,\psi^2)\right)(v,w) := \psi^1.v - \psi^2.w.$$
Then $v^1 \in I$ is equivalent to
$$(v_1, v_2) \notin \ker \rho(\tilde{L}_{N_2})|_{V^1_{N_2 - 1} \times 
V^1_{N_2 - 2}},$$
where $\ker \rho(\tilde{L}_{N_2})|_{V^1_{N_2 - 1} \times V^1_{N_2 - 2}}$
is the intersection of the kernels of all its elements.
We now proceed by proving
$$\dim \rho(\tilde{L}_{N_2})|_{V^1_{N_2 - 1} \times V^1_{N_2 - 2}}
= \dim \tilde{L}_{N_2}.$$
\noindent To this end we assume the existence of $0 \not= (\chi, \psi) \in 
\tilde{L}_{N_2}$, such that $\chi.v - \psi.w = 0 \;\forall
(v,w) \in V^1_{N_2 - 1} \times V^1_{N_2 - 2}$.
With the involution (\ref{invol}) it follows that
$\chi^{\dag}.v^{(0)}, \psi^{\dag}.v^{(0)} \in \mbox{Rad}(\skpi{.}{.})$. By
theorem \ref{embedding} $(\chi, \psi) = 0$, which contradicts
the assumption.
\par\noindent
We can now determine the dimension of the allowed parameter space 
$\tilde{\mathcal{V}}_{h^1,h^2} \subset V^1_{N_2-1} \times V^1_{N_2-2}$
for $(v_1,v_2)$.
With (\ref{dimexample}) and because of 
$\rho(\tilde{L}_{N_2})|_{V^1_{N_2 - 1} \times V^1_{N_2 - 2}} \subset
\Hom(V^1_{N_2 - 1} \times V^1_{N_2 - 2}, \CC)$, it is
\[\label{indbeginproof}\dim \tilde{\mathcal{V}}_{h^1,h^2} = 
\dim V^1_{N_2-1} + \dim V^1_{N_2-2} - \dim \tilde{L}_{N_2} = \p(N_2).\]
\par~\par\noindent
After having determined the allowed parameter space we now examine which
of these choices lead to isomorphic modules. Firstly, it is clear, that
any two vermalike staggered modules whose data $\{v_0,v_1,v_2\}$ only differ by
a nonzero scalar factor are isomorphic. We therefore fix the scaling of the
upper LWV by demanding that $v_0 = v^2 = \chi_1.v^1$. Now two vermalike 
staggered modules $M, M'$ are isomorphic if and only if there is a vector
$\tilde{v} \in V^1_{N_2}$ so that $v'_1 = v_1 + L_{-1} \tilde{v}$ and
$v'_2 = v_2 + L_{-2} \tilde{v}$. The space of all nonisomorphic vermalike
staggered modules with lowest weights $h^1$ and $h^2$ is given by
\[\mathcal{V}_{h^1,h^2} = \tilde{\mathcal{V}}_{h^1,h^2} / 
(L_{-1}, L_{-2}).V^1_{N_2}.\]
By theorem \ref{embedding} the dimension of $(L_{-1}, L_{-2}).V^1_{N_2}$
simply is
\[\dim V^1_{N_2} -1 = \p(N_2) -1.\]
Put together, this yields the resulting
\begin{lemma} \label{directsucc} 
With the notations of definition \ref{notation} the
space of all nonisomorphic staggered modules with lowest weights
$h^1$ and $h^2$ is a vector space $\mathcal{V}_{h^1, h^2}$ with
dimension $\dim \mathcal{V}_{h^1, h^2} = \p(N_2) - (\p(N_2) -1) = 1$.\qed
\end{lemma}
\noindent
The above moduli space is most naturally parametrized in the following way:
Obviously $\chi_1^\dagger.v^{(1)} = \alpha v^1$ with $\alpha \in \CC$.
$\alpha = 0$ is equivalent to $M$ belonging to the equivalence class
of $(v_0,v_1,v_2) = (\chi_1.v^1, 0,0)$:
Let $\{\psi_i.v^1,\;\psi_0 = \chi_1\}$ denote an orthogonal base of $V^1_{N_2}$
with respect to the Shapovalov form on $V^1$ with 
$\skpi{\psi_i.v^1}{\psi_j.v^1} = s_i \delta_{ij}$. Theorem \ref{embedding}
implies that $s_i = 0 \Leftrightarrow i = 0$. If $\alpha = 0$ then
$\tilde{v}^{(1)} := v^{(1)} - \sum_{k = 1}^{\p(N_2)-1} s_k^{-1} 
\psi_k.{\psi_k}^\dagger v^{(1)}$
also is an upper lowest weight vector with $L_0.\tilde{v}^{(1)} = \chi_1.v^1$
and $L_{-1}.\tilde{v}^{(1)} = L_{-2}.\tilde{v}^{(1)} = 0$. On the other
hand, due to $\chi_1.v^1$ belonging to the radical of $\skpi{.}{.}$, $\alpha$ 
does not depend on the choice of the representative $(v_1,v_2)$.
\par~\par\noindent
We will now extend this result to the general case $h^{(0)} = h^1$ and
$h^{(0)} = h^k, k\in \NN^{\ge 2}$. One first observes
\begin{lemma} \label{alwaysdistone}
Let $M$ be a vermalike staggered module with lowest weights $h^1$ and
$h^k, k\in \NN^{\ge 2}$. Then it is always possible to choose
an upper LWV so that $v_1, v_2 \in V^{k-1} \subset M^{(0)}$.
\end{lemma}
\begin{proof}
It suffices to show the following: If
$v_1, v_2 \in V^j, j < k-1$, then there is a choice of upper lowest weight
vector which yields $\tilde{v}_1, \tilde{v}_2 \in V^{j+1}$. Let
$\{\psi_i \chi_j.v^j, \xi_l.v^j; 0 \le i < \p(h^k-h^{j+1}) 
\le l < \p(h^k-h^j)\}$
denote an orthogonal base of $V^j_{h^k-h^j}$ with respect to the Shapovalov
form on $V^j$. As the Shapovalov form is nondegenerate on $V^j / V^{j+1}$,
one has
$\skpi{\psi_p \chi_j.v^j}{\psi_q \chi_j.v^j} = 
\skpi{\psi_p \chi_j.v^j}{\xi_r.v^j} = 0$ 
and $\skpi{\xi_r.v^j}{\xi_s.v^j} = s_r \delta_{rs},\; s_r \not= 0$.
As $v_1, v_2 \in V^j$, we have 
$(\psi_i \chi_j)^\dagger.v^{(1)} = \chi_j^\dagger 
({\psi_i}^\dagger.v^{(1)}) = 0$.
Let $\tilde{v}^{(1)} := v^{(1)} - \sum_{r = \p(h^k-h^{j+1})}^{\p(h^k-h^j)-1} 
s_r^{-1} \xi_r.{\xi_r}^\dagger v^{(1)}$.
Then by construction $\UU^-_{h^k-h^j}.\tilde{v}^{(1)} = 0$ and therefore
$\tilde{v}_1, \tilde{v}_2 \in \mbox{Rad}(\skpi{.}{.}) = V^{j+1}$.\qed
\end{proof}
\noindent We can now state our final result:
\begin{theorem} \label{staggmodspace}
With the notations of definition \ref{notation} the
space of all nonisomorphic staggered modules with lowest weights
$h^1$ and $h^k,k \in \NN^{\ge 2},$ is a vector space $\mathcal{V}_{h^1, h^k}$ 
with dimension $\dim \mathcal{V}_{h^1, h^k} = 1$.
\end{theorem}
\begin{proof}
By lemmata \ref{directsucc} and \ref{alwaysdistone} we have 
\[\dim \mathcal{V}_{h^1,h^k} \le 1.\label{beginineq}\] 
Let $v_1, v_2 \in V^{k-1}$ be data of a
staggered module with lowest weights $h^{k-1}, h^k$, which is not isomorphic
to the staggered module with $v_1 = v_2 = 0$.
The only thing which could prevent equality in (\ref{beginineq}) is the
existence of a vector $\tilde{v} \in V^1 \backslash V^{k-1}$ so that
$L_{-1} \tilde{v} = v_1$ and $L_{-2} \tilde{v} = v_2$. But then 
$\tilde{v}+V^{k+1}$ would be singular in $V^1 / V^{k-1}$.
By theorem \ref{embedding} it follows that $\tilde{v}+V^{k+1} = 0+V^{k+1}$
and therefore $\tilde{v} \in V^{k+1}$, which contradicts the assumption.\qed
\end{proof}
\begin{remark} By using so-called central series \cite{fefu3}, one can show the
restrictions of corollary \ref{weightrestrict} even for the more general
case of arbitrary extensions of a Verma module $V(h^{(0)},c)$ by another
Verma module $V(h^{(1)},c)$. Almost the whole preceding argumentation goes
through for the case of N-length 1, yielding two nonisomorphic modules --
one nontrivial extension with N-length 1 and the (trivial) direct sum of the
two modules:
\par\noindent
The vector space $\Ext(V(h^1,c),V(h^k,c))$ of nonequivalent exact sequences
$$0 \longrightarrow V(h^1,c) \stackrel{\beta}{\longrightarrow} M 
\stackrel{\gamma}{\longrightarrow} V(h^k,c) \longrightarrow 0,$$
where two sequences are equivalent if there is an \LL-isomorphism 
$\alpha: M \rightarrow M'$ such that the diagram
$$\begin{array}{ccccccccc}
0 & \longrightarrow & V(h^1,c) & \stackrel{\beta}{\longrightarrow} & M &
\stackrel{\gamma}{\longrightarrow} & V(h^k,c) & \longrightarrow & 0 \\
& & \Big\| & & \Big\downarrow{\scriptstyle\alpha} & & \Big\| & & \\
0 & \longrightarrow & V(h^1,c) & \stackrel{\beta'}{\longrightarrow} & M' &
\stackrel{\gamma'}{\longrightarrow} & V(h^k,c) & \longrightarrow & 0
\end{array}$$
commutes, has complex dimension 2. The vector space $\mathcal{V}_{h^1, h^k}$
from theorem \ref{staggmodspace} is the one-dimensional affine subspace of
$\Ext(V(h^1,c),V(h^k,c))$, where $L_0^n v^{(1)}$ is fixed to a nonzero value.
The linear subspace $\mathcal{V}$ parallel to $\mathcal{V}_{h^1, h^k}$ is 
exactly the subspace where $\nlength(M) = 1$.
\par\noindent
The space of all nonisomorphic modules $M$ is then
given by 
$\mathrm{P}(\Ext(V(h^1,c),V(h^k,c))) := \Ext(V(h^1,c),V(h^k,c)) / \sim$,
where two sequences are equivalent (`$\sim$'), if there is 
a number $\delta\in\CC^{\not=0}$ and an \LL-isomorphism 
$\alpha: M \rightarrow M'$, such that the diagram 
$$\begin{array}{ccccccccc}
0 & \longrightarrow & V(h^1,c) & \stackrel{\beta}{\longrightarrow} & M &
\stackrel{\gamma}{\longrightarrow} & V(h^k,c) & \longrightarrow & 0 \\
& & \Big\| & & \Big\downarrow{\scriptstyle\alpha} & & 
\Big\downarrow{\scriptstyle\delta\cdot id} & & \\
0 & \longrightarrow & V(h^1,c) & \stackrel{\beta'}{\longrightarrow} & M' &
\stackrel{\gamma'}{\longrightarrow} & V(h^k,c) & \longrightarrow & 0
\end{array}$$
commutes (this is a well defined equivalence relation on
$\Ext(V(h^1,c),V(h^k,c))$, since the old equivalence relation is just a
special case ($\delta = 1$) of the new one)\footnote{For a given vector
space $V$ the space $\mathrm{P}(V)$ is the union 
of the corresponding projective
space and one isolated point: $\mathrm{P}(V) = \PP(V)\,\dot{\cup}\,\{0\}$.}.
\par\noindent
$\mathrm{P}(\Ext(V(h^1,c),V(h^k,c)))$ splits as follows:
\begin{eqnarray*}\mathrm{P}(\Ext(V(h^1,c),V(h^k,c))) & = & 
\mathcal{V}_{h^1, h^k} \,\dot{\cup}\, \mathrm{P}(\mathcal{V})\\
& = & \mathcal{V}_{h^1, h^k} \,\dot{\cup}\, 
\{\mbox{\textup{extension~with $\nlength$ 1}}\} 
\,\dot{\cup}\, \{V(h^1,c) \oplus V(h^k,c)\}.
\end{eqnarray*}
\end{remark}

\section{The maximal preserving submodules of JVMs}
We are now prepared to explore the maximal preserving submodules
of JVMs. As a first approach we prove the following theorem:
\begin{theorem} \label{alwaysdistonesub}
Let $\tilde{V}(h^1, c)$ a JVM and notations as
in definition \ref{notation}. For any $n,m \in \NN, m > n$ there exists
a submodule $J \subset \tilde{V}(h^1, c)$ with
$J^{(0)} = V^n$ and $J^{(1)} \cong V^m$.
\end{theorem}
\begin{proof}
Let $\tilde{J} := \UU.\chi_{m-1} \ldots \chi_1 v^{(1)} + 
\tilde{V}(h^1, c)^{(0)}$ denote the staggered submodule of 
$\tilde{V}(h^1, c)$ with lowest weights $h^1$ and $h^m$.
According to lemma \ref{alwaysdistone} $\tilde{J}$ possesses a
staggered submodule $J \subset \tilde{J} \subset \tilde{V}(h^1, c)$ 
with lowest weights $h^{m-1}$ and $h^m$. All other cases are given
by the submodules $J + V^n$.\qed
\end{proof}
\noindent
By lemma \ref{firstclassify} and the above theorem \ref{alwaysdistonesub} 
a JVM $V$ with maximal preserving submodule $J$ falls 
into one of two classes: In any case $J^{(0)} = V^2$, but either
$J^{(1)} \cong V^3$ or $J^{(1)} \cong V^2$, i.e. there exists an 
embedding $V(h^2, c) \hookrightarrow V(h^1, c)$.
We now want to study the relationship between membership in one of
the above classes and the lowest weight of the module. The means to
do so are -- again -- provided by the Shapovalov form.
In order to simplify notation we first define a projector
$P: \UU_0 \rightarrow \UU_0$ by linear continuation of the following 
settings:\par\noindent 
For any monomial 
$u := L_{m_1}\ldots L_{m_p} L_0^{k_0} C^{k_c} L_{n_1} \ldots L_{n_q} \in \UU_0$
with $m_1 \ge \ldots \ge m_p > 0 > n_1 \ge \ldots \ge n_q$ let
$$P(u) = \left\{\begin{array}{ll} u & \mbox{if $p = q = 0$} \\   
0 & \mbox{otherwise.} \end{array}\right.$$
Clearly, $u.v = P(u).v$ on a singular vector $v$.
\begin{lemma} \label{doublezero}
With the notations of definition \ref{notation} the Kac determinant
$\det \skpi{.}{.}_{N_2}(h)$ possesses a double zero at $h = h^1$ if
and only if $P(\chi_1^\dagger \chi_1) = \chi_1^* (L_0-h^1)^2$ with
$\chi_1^* \in P(\UU_0)$.
\end{lemma}
\begin{proof}
Let $\{\psi_0.v^1 := \chi_1.v^1, \psi_i.v^1; 1 \le i < \p(N_2)\}$ be an 
orthogonal base of $V(h^1,c)$ with respect to $\skpi{.}{.}_{N_2}(h^1)$.
We define the polynomials
$$p_{i,j}(h) := \skpi{\psi_i v_h}{\psi_j v_h}$$
where $v_h$ is the lowest weight vector of $V(h,c)$.
Then obviously
\[\det \skpi{.}{.}_{N_2}(h) = \det \left(\begin{array}{ccc}
p_{0,0} & \cdots & p_{0,\p(N_2)-1} \\
\vdots & & \vdots \\
p_{\p(N_2)-1,0} & \cdots & p_{\p(N_2)-1,\p(N_2)-1}
\end{array}\right).\]
As the first row and the first column vanish for $h = h^1$,
the polynomials $p_{0,i}$ and $p_{i,0}$ have a common factor
$(h-h^1)$.
If we expand the determinant by the first column, we see that
$$\det \skpi{.}{.}_{N_2}(h) = p_{0,0}(h) p^*(h) + (h-h^1)^2 \tilde{p}(h)$$
where $p^*(h^1) \not= 0$ because of theorem \ref{embedding}.
Hence, the determinant possesses only a single zero at $h= h^1$
if and only if $p_{0,0}$ possesses only a single zero, which proves
the assertion.\qed
\end{proof}
\noindent
By close examination of the Kac determinant formula (\ref{kacdetformula}), 
the knowledge, that
each pair $(r,s)$ with $h_{r,s}=h$ corresponds to a singular vector on
level $rs$ \cite{fefu1}, and by careful study of the symmetries of $h_{r,s}$
one obtains the additional result:
\begin{lemma} \label{whendoublezero}
As before, let $c = c_{1,q}$. Let 
$h_n := \frac{n^2-(q-1)^2}{4 q}$.
$\det \skpi{.}{.}_{N_2}(h)$ possesses a double zero at $h=h^1$
if and only if $h^1 = h_n$ where $n \not= 0$ is a multiple of $q$.
\end{lemma}
\begin{theorem} \label{whenjordan}
With notations as in definition \ref{notation} let 
$\tilde{V}(h_n,c)$ be a JVM with lowest weight
$h_n = \frac{n^2-(q-1)^2}{4 q}$. The maximal preserving submodule
$J \subset \tilde{V}(h_n,c)$ is a JVM if and only if
$n \not= 0$ is a multiple of $q$.
\end{theorem}
\begin{proof}
The maximal preserving submodule $J$ is a JVM if and
only if
$$\exists \tilde{v} \in V^1_{N_2}:\;\UU^-.(\chi_1.v^{(1)} + \tilde{v}) = 0.$$
This is equivalent to
$$\chi_1^\dagger \chi_1.v^{(1)} = 0,$$
as due to $\chi_1.v^1$ belonging to the radical of the Shapovalov form
$$\nexists \tilde{v} \in V^1_{N_2}:\; \chi_1^\dagger \tilde{v} \not= 0.$$
If $\chi_1^\dagger \chi_1.v^{(1)} = 0,$ a suitable choice is given by
$$\tilde{v} = -\sum_{i\not=0} s_i^{-1} \psi_i \psi_i^\dagger \chi_1 v^{(1)},$$
where the $\psi_i$ and $s_i$ are defined as in the proof of lemma
\ref{doublezero}. With lemmata \ref{doublezero} and \ref{whendoublezero} 
the assertion follows.\qed
\end{proof}
\begin{remark}
The proof of theorem \ref{whenjordan} uses special properties of the 
rank 2 case, namely the vanishing of $(L_0^n)^2$ on the upper lowest weight
vector. Therefore, the possibility to embed a JVM into another, also is a
genuine property of the rank 2 case.
\end{remark}
\subsection{Characters}
The character of an \LL-module $V$ is defined as
$$\chi^{}_V(q) := \tr_V q^{L_0 - \frac{c}{24}} = 
q^{- \frac{c}{24}} \sum_h q^h \dim \eigspace(L_0^d,h).$$
By lemma \ref{vermapbwbase} the character of a Verma module therefore is
given by
$$\chi^{}_{V(h,c)} = q^{h - \frac{c}{24}} \sum_{k=0}^\infty \p(k) q^k =
\frac{q^{\frac{1-c}{24}}}{\eta(q)} q^h,$$
where $\eta$ is the Dedekind $\eta$-function $\eta(q) = q^\frac{1}{24} 
\prod_{n\in\NN}(1-q^n)$.
\begin{corr}\label{lirrchar}
With the notations of definition \ref{notation} let $M(h^1,c)$ be the
irreducible \LL-module with lowest weight $h^1$. Its character is
given by
$$\chi^{}_{M(h^1,c)} = \chi^{}_{(V(h^1,c) / V(h^2,c))}
= \frac{q^{\frac{1-c}{24}}}{\eta(q)} (q^{h^1}-q^{h^2}).$$
\end{corr}
\noindent
With the results of theorems \ref{alwaysdistonesub} and \ref{whenjordan}
we obtain the
\begin{corr}
With the notations of definition \ref{notation} let $\tilde{M}(h,c)$ be 
a minimal JLWM. Its character is given by one of the following three formulas:
If $\nexists r,s\in\NN:\;h= h_{r,s}$, we have
$$\chi^{}_{\tilde{M}(h,c)} = 2 \frac{q^{\frac{1-c}{24}}}{\eta(q)} q^h.$$
Else, $h = \frac{n^2-(q-1)^2}{4 q}$ with $n \in \NN^0$.
If $n \not= 0$ and $n$ is a multiple of $q$, then
$$\chi^{}_{\tilde{M}(h,c)} = 2 \frac{q^{\frac{1-c}{24}}}{\eta(q)} 
(q^{h^1} - q^{h^2}).$$
In all other cases, 
$$\chi^{}_{\tilde{M}(h,c)} = \frac{q^{\frac{1-c}{24}}}{\eta(q)} 
(2 q^{h^1} - q^{h^2} - q^{h^3}).$$ 
\end{corr}
\begin{corr}
The character of a vermalike staggered module $V$ with lowest weights
$h^1$ and $h^k$ is given by
$$\chi^{}_{V} = \frac{q^{\frac{1-c}{24}}}{\eta(q)} 
(q^{h^1}+q^{h^k}).$$
If its characteristic parameter (see subsection \ref{modspaces})
$\alpha = 0$, the character of the corresponding minimal
staggered module $\tilde{V}$ is
$$\chi_{\tilde{V}} = \frac{q^{\frac{1-c}{24}}}{\eta(q)} 
(q^{h^1}+q^{h^k}-2q^{h^{k+1}}).$$
In all other cases it is given by
$$\chi_{\tilde{V}} = \frac{q^{\frac{1-c}{24}}}{\eta(q)} 
(q^{h^1}+q^{h^k}-q^{h^{k+1}}-q^{h^{k+2}}).$$
\end{corr}

\section{Embeddings}
The structure of minimal JVMs of rank 2 was completely resolved
by theorems \ref{alwaysdistonesub} and \ref{whenjordan}.
This only involved the question, whether a JVM can be
embedded into another as a \emph{maximal} preserving submodule.
\par\noindent
If, with the notations of definition \ref{notation} and
theorem \ref{whenjordan},
$h = h_n$ with $n \not= 0$ and $n$ a multiple of $q$,
this question can easily be answered: As the lowest weight
of the maximal preserving submodule of $\tilde{V}(h,c)$
again fulfills the above condition (see theorem \ref{embedding}),
the complete embedding structure is given by
$$\begin{array}{ccccccccccccccc}
 & & & & & & & & & & & & \vdots & & \\
 & & & & & & & & & & & & \downarrow & & \\
 & & & & & & & & J_{5,5} & \leftarrow & \cdots &
        \leftarrow & J_{5,\infty} & = & V^5\\
 & & & & & & & & \downarrow & & & & \downarrow & & \\
 & & & & & & J_{4,4} & \leftarrow & J_{4,5} & \leftarrow & \cdots &
        \leftarrow & J_{4,\infty} & = & V^4\\
 & & & & & & \downarrow & & \downarrow & & & & \downarrow & & \\
 & & & & J_{3,3} & \leftarrow & J_{3,4} & \leftarrow & J_{3,5} & \leftarrow &
        \cdots & \leftarrow & J_{3,\infty} & = & V^3\\
 & & & & \downarrow & & \downarrow & & \downarrow & & & & \downarrow & & \\
 & & J_{2,2} & \leftarrow & J_{2,3} & \leftarrow & J_{2,4} & \leftarrow &
        J_{2,5} & \leftarrow & \cdots & \leftarrow & J_{2,\infty} & = & V^2\\
 & & \downarrow& & \downarrow & & \downarrow & & \downarrow & & & &
        \downarrow & & \\
\tilde{V}(h,c) & \leftarrow & J_{1,2} & \leftarrow & J_{1,3} & \leftarrow & 
        J_{1,4} & \leftarrow & J_{1,5} & \leftarrow & \cdots & \leftarrow &
        J_{1,\infty} & = & V^1,\\
\end{array}$$
\noindent where $J_{k,l}$ denotes a staggered submodule with lowest weights
$h^k$ and $h^l$.
In general, one always has an embedding structure of the form
$$\begin{array}{ccccccccccccccc}
 & & & & & & & & & & & & \vdots & & \\
 & & & & & & & & & & & & \downarrow & & \\
 & & & & & & & & (J_{5,5}) & \leftarrow & \cdots &
        \leftarrow & J_{5,\infty} & = & V^5\\
 & & & & & & & & \downarrow & & & & \downarrow & & \\
 & & & & & & (J_{4,4}) & \leftarrow & J_{4,5} & \leftarrow & \cdots &
        \leftarrow & J_{4,\infty} & = & V^4\\
 & & & & & & \downarrow & & \downarrow & & & & \downarrow & & \\
 & & & & (J_{3,3}) & \leftarrow & J_{3,4} & \leftarrow & J_{3,5} & \leftarrow &
        \cdots & \leftarrow & J_{3,\infty} & = & V^3\\
 & & & & \downarrow & & \downarrow & & \downarrow & & & & \downarrow & & \\
 & & (J_{2,2}) & \leftarrow & J_{2,3} & \leftarrow & J_{2,4} & \leftarrow &
        J_{2,5} & \leftarrow & \cdots & \leftarrow & J_{2,\infty} & = & V^2\\
 & & \downarrow& & \downarrow & & \downarrow & & \downarrow & & & &
        \downarrow & & \\
\tilde{V}(h,c) & \leftarrow & J_{1,2} & \leftarrow & J_{1,3} & \leftarrow & 
        J_{1,4} & \leftarrow & J_{1,5} & \leftarrow & \cdots & \leftarrow &
        J_{1,\infty} & = & V^1,\\
\end{array}$$
\noindent
where the submodules in brackets may or may not exist.
Unfortunately at the time being we cannot in general answer the
question, which of these modules exist for given lowest weight
$h^1$. Therefore we restrict ourselves to listing the restrictions
we know:
\begin{enumerate}
\item If $h^1 = h_n$ with $n \not= 0$ and $n$ a multiple of $q$, all
of the $J_{n,n}$ exist.
\item If $h^1 = h_0$, a submodule $J_{2,2}$ does not exist. If
the submodule $J_{N,N}$ exists for any $N\in\NN^{\ge 3}$, then all 
submodules $J_{n,n}$ with $n \ge N$ also exist.
\item If $h^1 = h_n$ with $n/q \not\in \NN^0$, a submodule $J_{2,2}$
does not exist. If, for any $n \in \NN$, the submodule $J_{n,n}$ exists,
then no submodule $J_{n+1,n+1}$ can exist.
\end{enumerate}

\section{Rational models}
It is a natural question, whether there are rational conformal
field theories at values of the central charge from the
logarithmic series. Of course, the notion of rationality here has to be 
broadened to include indecomposable representations of the underlying 
\WW-algebra (for a short and exact definition of \WW-algebra see e.g. 
\cite{wdef}).
In the mathematical literature one often defines
a rational theory to be decomposable into finitely many \emph{irreducible} 
representations, which close under fusion -- this clearly cannot be
the case for the above central charges. We will later see, 
in which way the usual definition of
rationality has to be broadened to include logarithmic models 
(definition \ref{rationaldef}). 
\par\noindent 
While it is not possible to find rational theories with
respect to the Virasoro algebra, at least for some of the central
charges in the logarithmic series one indeed finds
models of larger \WW-algebras, which are rational in this 
slightly broadened sense.
\par\noindent
The right candidates for such rational theories were already identified
by M.~Flohr \cite{flohr1,flohr2}. He found that the characters of the
known representations of a suitably chosen \WW-algebra span finite
dimensional representations of the modular group $SL(2,\ZZ)$.
The problem with these representations was that they necessarily include
``characters'' with logarithmic terms in $q$, which, at least with the usual
definition of a character, cannot occur. In fact, one can find
finite dimensional representations of the modular group without the inclusion
of logarithmic ``characters'' if one does not start with the set of usual
lowest weight representations of the algebra but rather with some
suitable extensions thereof.
\par\noindent
To see this, we will now study the simplest example of such a rational 
logarithmic model. It is based on the triplet algebra $\WW(2,3^3)$ at $c=-2$.
This algebra was found by H.~Kausch \cite{tripletalg} and is spanned by the 
modes of the Virasoro field and three additional primary fields of conformal
weight three. The algebra is given by the commutation relations
\begin{myeqnarraylabel}{wcomm} 
\left[ L_m,L_n \right] & = & (n-m) L_{m+n} - \frac{1}{6} (n^3-n) 
\delta_{n+m,0},\\
\left[ L_m,W^a_n \right] & = & (n-2m)W^{a}_{m+n},\\
\left[ W^{a}_m,W^{b}_n \right] & = & g^{ab}\left(
    4\,p_{334}(m,n) \Lambda_{m+n} + 3\,p_{335}(m,n) L_{m+n} \right.\nonumber\\
& &  \left. - \binom{n+2}{5} \delta_{m+n,0}\right)\nonumber\\
& &  + f_c^{ab}\left(5\,p_{333}(m,n) W^c_{m+n} + 
\frac{12}{5}\Omega^c_{m+n}\right)
\end{myeqnarraylabel}
\noindent
where $a,b \in \{1,2,3\}$, 
$\Lambda := \Nop(L,L) = N(L,L)-3/10\,\partial^2L$ and 
$\Omega^a := \Nop(W^a,L) = N(W^a,L) -3/14\,\partial^2W^a$
are quasiprimary normal ordered fields (for
notational conventions see the appendix). 
The $p_{ijk}$ are universal polynomials:\\
$p_{332}(m,n) = \frac{n-m}{60}(2m^2+2n^2-mn-8)$,
$p_{333}(m,n) = \frac{1}{14}(2m^2+2n^2-3mn-4)$
and $p_{334}(m,n) = \frac{n-m}{2}$.
$f^{ab}_c$ and $g^{ab}$ are the structure
constants and standard symmetric bilinear form of $su(2)$ 
(the latter is half the Killing form on $su(2)$, 
i.e. in the standard base with 
$f^{ab}_c = i \eps_{abc}$ one has $g^{ab} = \delta_{a,b}$).
\par\noindent
Before we study the above-mentioned rational model, we must first slightly 
generalize our definitions from sections \ref{lowestsect} and
\ref{generalcase}. For the sake of notational simplicity we will
concentrate on the above defined \WW-algebra $\WWW$ at $c = -2$. 
The generalizations to other \WW-algebras will be obvious. 
\par\noindent
We first remark that irreducible modules
of a given \WW-algebra are not necessarily lowest weight modules.
The role of lowest weight modules will be played by a slightly broader
class of modules, which are based on irreducible representations of 
the subalgebra of zero modes (in the case of $\WWW$ the subalgebra
generated by $\{L_0,W_0^1,W_0^2,W_0^3\}$).

\begin{define}
Let $\cmodw$ be the category of \WWW-modules, which as \LL-modules
belong to $\cmodl$. Its objects will from now on simply be called
\WWW-modules.
\end{define} 

\begin{lemma}
One has $[L_0^n, W^a_k] = 0$. Furthermore
\[W^a_k: \eigspace(L_0^d, h) \rightarrow \eigspace(L_0^d, h+k).\]
For an indecomposable \WWW-module $V$ one therefore has
\[V = \bigoplus\limits_{n=0}^{\infty} \eigspace(L_0^d, h_{min}+n).\]
\end{lemma}
\begin{proof}
Analogous to subsection \ref{gradfilt}.\qed
\end{proof}

\begin{define}
A $\WWW$-module $M \in \cmodw$ is called \textbf{generalized lowest weight 
module (GLWM)} if there is a linear subspace $M_0 \subset M$ such that
\begin{enumerate}
\item $\forall n\in\NN, v\in M_0, a\in\{1,2,3\}: L_{-n} v = W^a_{-n} v = 0$,
\item $M_0$ is an irreducible $\WW_0$-module,
\item $M = \UW.M_0$,
\end{enumerate}
where $\UW$ denotes the universal enveloping algebra of $\WWW$. If
$\dim M_0 = 1$, $M$ is called \textbf{lowest weight module} or
\textbf{singlet module}. For $\dim M_0 = 2$, $M$ is called
\textbf{doublet module}.
\end{define}

\begin{lemdef} \label{lscalar}
Let $M \in \cmodw$ be a generalized lowest weight module. Then
$\exists h\in\CC$ such that $\forall v \in M_0: L_0.v = h v$. $h$ is called
\textbf{lowest weight} of the module and the elements of a base
of the \textbf{lowest weight space} $M_0$ are called
\textbf{lowest weight vectors}.
\end{lemdef}
\begin{proof} From (\ref{wcomm}) we have $[L_0,W^a_0] = 0$ (this is true for
arbitrary primary fields $W^a$). With Schur's lemma the assertion
follows.\qed
\end{proof}

\begin{lemma}\label{indecdiagirr}
Let $M \in \cmodw$ be indecomposable and $L_0 M_0 = h M_0, h\in\CC$. 
Then the $\WW_0$-module $M_0$ is not only indecomposable, but
irreducible.
\end{lemma}
\begin{proof}
Because of $W^a_{-n} M_0 = L_{-n} M_0 = 0$, $n\in\NN$, for $v\in M_0$ one
has $\UW v \cap M_0 = \UU(\WW_0) v$. Hence, the $\WW_0$-module $M_0$
is indecomposable.
\par\noindent
For $v\in M_0$ one easily calculates
$$\Omega^a_0 v = \left(h+\frac{3}{7}\right) W^a_0 v.$$
With (\ref{wcomm}) it is clear, that the representation of 
$\langle W^1_0,W^2_0,W^3_0 \rangle$ on $M_0$ is just a representation 
of $su(2)$. $su(2)$ being semisimple, its finite 
dimensional representations are completely reducible by Weyl's
theorem (see e.g. \cite{humph}). 
Therefore, $M_0$ is both indecomposable and completely reducible,
hence irreducible.\qed
\end{proof}

\begin{lemma}
Let $V \in \cmodw$ be a \WWW-module. Then there exists a submodule
$M \subset V$ which is a generalized lowest weight module.
\end{lemma}
\begin{proof}
Without loss of generality suppose $V$ to be indecomposable.
Analogously to the proof of lemma \ref{thereislwm}, it
possesses a submodule $M$ with $L_0 M_0 = h M_0, h\in\CC$.
$M$ can also be chosen to be indecomposable. With lemma
\ref{indecdiagirr} $M_0$ then is an irreducible $\WW_0$-module
and $\UW M_0 \subset V$ is a GLWM.\qed
\end{proof}
\noindent
Using the above lemma, we may define the \textbf{lowest weight length}
of a \WWW-module analogously to definition \ref{lwlength}, if we
substitute \emph{lowest weight module} by \emph{generalized lowest
weight module}. The \textbf{length} and \textbf{nilpotency length} of a
\WWW-module are defined analogously to definitions \ref{lengthdefine} and
\ref{lwlength}.

As the analogues to staggered modules and JLWMs in 
the pure Virasoro case we define

\begin{define}
An indecomposable \WWW-module $M\in\cmodw$ is called \textbf{staggered module},
if $\nlength(M) = \llength(M) = N \in \NN^{\ge 2}$. The number $N$ is called
its \textbf{rank}. 
\end{define}

\begin{define}
A staggered \WWW-module $M$ of rank $N$ is called 
\textbf{Jordan lowest weight module} if $M = \UW. M_0$ and
$\forall v\in M_0:L_0 v = h v$.
\end{define}

\begin{define}
A staggered \WWW-module $M$ of rank 2 is called \textbf{strictly staggered},
if it is not a JLWM.
\end{define}

\noindent As we will see later, it also becomes necessary to introduce yet 
another class of modules:

\begin{define} \label{genstaggdef}
An indecomposable \WWW-module $M$ with $\nlength(M) = 2$ is called 
\textbf{generalized staggered module} if $M^{(1)}$ is a GLWM.
\end{define}

\subsection{Null field relations}
On first sight the above definitions seem to admit a
much larger class of modules than in the Virasoro case.
This is in fact not true, as modules of a given
\WW-algebra must meet some restrictions which do not occur in the
pure Virasoro case.
\par\noindent
In general the algebra is only consistent (fulfills the Jacobi
identities) due to certain null states in the vacuum representation.
The existence of these null states,
corresponding to so-called \emph{null fields}, i.e. fields with
vanishing two-point functions, poses additional restrictions
on representations of the algebra, namely the vanishing
of these null fields \cite{annilideal}.
The algebra $\WWW$ is only associative due to the following 
null states in the vacuum representation:
\begin{myeqnarray}
A^a & = & \left(2 L_3W^a_3-\frac{4}{3}L_2W^a_4 + W^a_6\right) |0\rangle \\
B^{ab} & = & W^b_3W^a_3 |0\rangle - g^{ab}\left(\frac{8}{9}L^3_2 + 
\frac{19}{36}L^2_3 + \frac{14}{9}L_4L_2 - \frac{16}{9}L_6\right) |0\rangle 
\nonumber\\
& & - f^{ab}_c \left(-2L_2W^c_4 + \frac{5}{4}W^c_6\right) |0\rangle.
\end{myeqnarray}
\begin{define}
Any \WWW-module $M$ fulfills
\[\forall v\in M,k\in\ZZ: \AAA^a_k v = \BBB^{ab}_k v = 0,
\label{nullrelation}\]
where $\AAA^a$ and $\BBB^{ab}$ denote the fields corresponding to
the null vectors $A^a$ and $B^{ab}$, respectively (see the appendix
for details). The property (\ref{nullrelation}) is called
\textbf{admissibility}, the module \textbf{admissible}.
\end{define}

\subsection{Generalized lowest weight modules}
\label{genlowsect}
We now want to study, which admissible generalized lowest
weight modules can exist. Admissible modules must be
annihilated by the null modes of $\AAA^a$ and $\BBB^{ab}$, respectively
-- in particular their lowest weight spaces must be annihilated,
where the action of the zero modes is especially easy to calculate. 
For $v \in M_0$ one obtains
\[\BBB^{ab} v =
\left(W_0^a W_0^b - g^{ab}\frac{1}{9}L_0^2 (8L_0+1) 
- f^{ab}_c\frac{1}{5}(6L_0-1)W_0^c\right) v = 0.\label{firstwrestrict}\]
The relation $\AAA^a v = 0$ is satisfied identically.
Further restrictions are obtained if one examines higher modes of the null 
fields. The study of the equations $W_{-1}^a \BBB^{bc}_1 v$, together
with \refeq{firstwrestrict}, after some lengthy but straightforward algebra
yields the result
$$W_0^a(8L_0-3)(L_0-1) v = 0,$$
which after multiplication by $W_0^a$ together with (\ref{firstwrestrict})
forces
\[L_0^2(8L_0 + 1)(8L_0-3)(L_0-1) v = 0.\label{wweightrestrict}\]
This restricts the lowest weights of generalized lowest weight modules to
$$h \in \left\{-\frac{1}{8}, 0, \frac{3}{8}, 1\right\}.$$
We now have to determine, which irreducible representations of
the zero mode algebra correspond to these values of $h$:
By redefining 
\[\tilde{W}_0^a := \frac{5}{2 (6h-1)} W_0^a,\label{sutwoscaling}\] 
one obtains on $M_0$ the $su(2)$-algebra
$$[\tilde{W}_0^a,\tilde{W}_0^b] = f^{ab}_c \tilde{W}_0^c.$$
Its Casimir operator $\sum\limits_{a,b} g_{ab} \tilde{W}_0^a \tilde{W}_0^b$ 
can then easily be evaluated using (\ref{firstwrestrict}):
\begin{center}
\renewcommand{\arraystretch}{1.5}
\begin{tabular}{|l||*{4}{c|}}
\hline
$h$ & $-\frac{1}{8}$ & $0$ & $\frac{3}{8}$ & $1$ \\
\hline
$\sum\limits_{a,b} g_{ab} \tilde{W}_0^a \tilde{W}_0^b$ & $0$ & $0$ & 
    $\frac{3}{4}$ & $\frac{3}{4}$ \\
\hline
\multicolumn{5}{c}{table \ref{genlowsect}.1: Admissible GLWMs}
\end{tabular}
\end{center}
\noindent From this result we conclude, that there exist at most
four inequivalent 
admissible irreducible \WWW-modules\footnote{Uniqueness is proven analogously
to theorem \ref{vermapbwbase}/corollary \ref{uniqueirrlwm}.}: 
Two singlet modules at $h=-\frac{1}{8},0$ 
(from now on called $V^{\WW}_{-1/8}$ and $V^{\WW}_{0}$) and two
doublet modules at $h = \frac{3}{8}, 1$ ($V^{\WW}_{3/8}$ and $V^{\WW}_{1}$). 
The singlet module at $h = 0$
is of course just the vacuum representation. 
All four modules have been obtained in
\cite{modconstruct} using a free field construction.
\par\noindent
The question, which \emph{reducible} admissible generalized lowest weight 
modules might exist, is also easily answered, as their maximal proper 
submodules must again be admissible.
In particular, the lowest weight space of a maximal proper submodule
must fulfill \refeq{wweightrestrict}, which only allows a reducible 
generalized LWM at $h = 0$.
In fact, such a module does exist: The \WWW-Verma module 
$\UW / \langle L_i,W^a_i; i\le0,a=1,2,3\rangle$ with lowest weight $0$
possesses two generalized lowest weight submodules with lowest weight $1$ 
(doublet modules).
It cannot possess any nontrivial submodules with trivial intersection
with the first two levels, since such a submodule cannot be admissible.
We denote this module by $\tilde{V}^{\WW}_{0}$.
\par\noindent
For future convenience we will now fix a choice of base for $su(2)$ to a 
Cartan-Weyl base $\{l^+,l^-,l^0\}$, such that the nonvanishing structure 
constants are given
by $f^{+-}_0 = - f^{-+}_0 = 2$, $f^{0\pm}_{\pm} = - f^{\pm0}_{\pm} = \pm1$
and the nonvanishing coefficients of the standard bilinear form become
$g^{00} = 1$, $g^{\pm\mp} = 2$. The representation matrices in the 
two-dimensional irreducible representation of $su(2)$ then, with a suitably
chosen base $\{v^+,v^-\}$, are given by 
$$\tau^+ := \left(\begin{array}{cc}0 & 1 \\ 0 & 0\end{array}\right),\;
\tau^- := \left(\begin{array}{cc}0 & 0 \\ 1 & 0\end{array}\right),\;
\tau^0:=\frac{1}{2}\left(\begin{array}{cc}1 & 0 \\ 0 & -1\end{array}\right).$$
With $v$ the lowest weight vector of the above reducible GLWM, a choice of 
lowest weight vectors for the two $V^{\WW}_{1}$ submodules is then given by
\begin{myeqnarray}
v_1^+ := \psi_1^+ v, & \; & \psi_1^+ := W_1^+\\
v_1^- := \psi_1^- v, & \; & \psi_1^- := \left(\frac{1}{2}L_1 - W_1^0\right)
\end{myeqnarray}
\noindent
and
\begin{myeqnarray}
v_2^+ := \psi_2^+ v, & \; & \psi_2^+ := \left(\frac{1}{2}L_1 + W_1^0\right)\\
v_2^- := \psi_2^- v, & \; & \psi_2^- := W_1^-.
\end{myeqnarray}
The above choice of base of course is somewhat arbitrary, but will prove
useful later.
In agreement with (\ref{sutwoscaling}) one calculates
\[W^a_0 v_i^b = 2 \tau^a v_i^b,\;L_0 v_i^b = v_i^b.\]

\subsection{Jordan lowest weight modules}
\label{wjordan}
\noindent
Admissibility poses even stronger restrictions on JLWMs than on GLWMs.
For the derivation of \refeq{wweightrestrict} we only used, that
$L_{-n}$ and $W^a_{-n}$ annihilate the lowest weight space. Therefore,
(\ref{wweightrestrict}) is also valid on the lowest weight space of an
admissible JLWM. Hence, the only admissible JLWMs of $\WWW$ have
lowest weight $0$ and nilpotency length $2$. 
\par\noindent
One further calculates,
that there is no JLWM with either both stages isomorphic to
$V^\WW_0$ or both stages isomorphic to $\tilde{V}^\WW_0$ (calculations up to 
level 2 are sufficient to prove this). 
Therefore, there exists \emph{only one} \WWW-JLWM
(denoted by $V^{\WW*}_0$) and
$(V^{\WW*}_0)^{(0)} = V^\WW_0$, 
$(V^{\WW*}_0)^{(1)} = \tilde{V}^\WW_0$.
\par\noindent

\subsection{Staggered modules}
\label{wstaggered}
\noindent
Equation (\ref{wweightrestrict}) only allows strictly
staggered modules with lowest weights 0 and 1.
Now assume the existence of such a staggered module and
let $v^{(0)}$, $v^{(1)+}$ and $v^{(1)-}$ its lower and upper LWVs.
Obviously, $L_0^n v^{(1)\pm}$ and $(W^a_0 - 2\tau^a) v^{(1)\pm}$ must be
singular (all vectors on the first level of the lower module are) and
therefore
\begin{eqnarray}
0 & = & W_0^a L_{-1} v^{(1)\pm} = 
[W_0^a, L_{-1}] v^{(1)\pm} + L_{-1} W_0^a v^{(1)\pm}\nonumber\\
& = & -2W_{-1}^a v^{(1)\pm} + L_{-1} W_0^a v^{(1)\pm} = 
-2W_{-1}^a v^{(1)\pm} + 2 L_{-1} \tau^a v^{(1)\pm}\\
& \Longrightarrow & W_{-1}^a v^{(1)\pm} = \frac{1}{2} L_{-1} W_0^a v^{(1)\pm}
 = L_{-1} \tau^a v^{(1)\pm}.\label{ldeterminesall}
\end{eqnarray}
Hence, the action of $L_{-1}$ on $v^{(1)\pm}$ completely determines
the action of $W^a_{-1}$ on $v^{(1)\pm}$.
\par\noindent
We now must check, whether such a module can be admissible.
To this end we must study the restrictions
coming from the null fields $\AAA^a$, $\BBB^{ab}$, where we now
must take into account all field modes $W^a_n, L_n$ with $n \ge -1$.
If one evaluates the constraint from $\AAA^a_0 v^{(1)\pm}$, one finds
that
\[W^a_1 L_{-1} v^{(1)\pm} = L_1 W^a_{-1} v^{(1)\pm}.\label{lwcommute}\]
Therefore, either $L_{-1} v^{(1)\pm} = W^a_{-1} v^{(1)\pm} = 0$ or
$L_1 v^{(0)} = \lambda W_1^a v^{(0)}, \lambda \in\CC$.
The first case is not allowed, since it would mean the existence
of an admissible JLWM with lowest weight 1 (c.f. section \ref{wjordan}). 
The second implies, since there is no \emph{singlet} module at 
$h = 1$, that 
\[L_1 v^{(0)} = W_1^a v^{(0)} = 0.\label{extendmeanszero}\]
Hence, $L_0^n v^{(1)\pm} = 0$, which is not allowed for staggered modules.

\subsection{Generalized staggered modules}
\label{wgenstagg}
\noindent
While a strictly staggered \WWW-module does not exist, we may
still expect to find generalized staggered modules as defined in
definition \ref{genstaggdef}. Assume the existence of such a module $M$, 
which is not a staggered module.
As before, equation (\ref{wweightrestrict}) forces 
$M^{(1)} = V^\WW_1$. We further assume $M$ to be minimal in the sense, that it
does not contain any proper submodule, that also is a
generalized staggered module.
\par\noindent
Right from the beginning we know, that in
the decomposition of the lower module $M^{(0)}$ into irreducible
modules only the modules $V^\WW_0$ and $V^\WW_1$ can occur.
In addition we know, that in the above decomposition the module
$V^\WW_0$ \emph{must} occur, because there is no admissible JLWM with
lowest weight $1$. Therefore we know that
$L_0 M_0 = 0$. From subsection \ref{genlowsect} we further know, that
$W_0^a M_0 = 0$.
We will now successively study the further restrictions on 
the lower module posed by admissibility.
\par\noindent
Let us choose representatives $v^{(1)+}, v^{(1)-} \in M_1$ of the lowest weight
vectors of $M^{(1)}$. Let $\tilde{M}^{(0)} := 
\UW \langle L_{-1} v^{(1)\pm}, W^a_{-1} v^{(1)\pm} \rangle$.
Obviously, $\tilde{M}^{(0)}$ is a submodule of $M^{(0)}$.
In fact, $\tilde{M}^{(0)} = M^{(0)}$: Otherwise $M / \tilde{M}^{(0)}$
would either contain an admissible JLWM with lowest weight $1$ as a submodule,
or it would do so after modding out $\UW.\eigspace(L_0,0)$.
\par\noindent
With \refeq{ldeterminesall} we find, that $\dim M^{(0)}_0$ can be at most 2.
As there are no strictly staggered modules of $\WWW$, we have
$\dim M^{(0)}_0 = 2$. A base of $M^{(0)}_0$
is given by the two vectors
\[v^{(0)\pm} := L_{-1} v^{(1)\pm}.\]
Equation (\ref{lwcommute}) then implies that $\dim M^{(0)}_1 = 2$ and
a base thereof is given by
\[\tilde{v}^{(0)\pm} := L_1 v^{(0)\pm} = L_1 L_{-1} v^{(1)\pm}.\]
More precisely, using \refeq{lwcommute} and the commutator relations
(\ref{wcomm}), one finds
\[W_1^a L_{-1} v^{(1)\pm} = L_1 W_{-1}^a v^{(1)\pm} = 
-\frac{1}{2} L_1 L_{-1} W^a_0 v^{(1)\pm},\label{wltollw}\]
and therefore
\begin{myeqnarray}
\psi_1^\pm v^{(0)+} & = & 0,\\
\psi_2^\pm v^{(0)-} & = & 0,\\
\psi_2^\pm v^{(0)+} & = & \psi_1^\pm v^{(0)-} = \tilde{v}^{(0)\pm}.
\end{myeqnarray}
\noindent
The lower module therefore is isomorphic to
$$\left(\tilde{V}^\WW_0 \oplus \tilde{V}^\WW_0\right) / A,$$
where 
$$A := \UW \psi_1^\pm (v,0) + \UW \psi_2^\pm (0,v) + 
\UW (\psi_2^\pm v, -\psi_1^\pm v)$$
and $v$ is a LWV of $\tilde{V}^\WW_0$.
\par~\par\noindent
Using $\BBB^{ab}_0 v^{(1)\pm} = 0$, $W^a_{-1} \BBB^{cd} v^{(1)\pm} = 0$,
$(L_0^n)^2 = 0$ and $L_0^n M^{(0)} = 0$, after some rather lengthy
algebra one finds
$$14 W_1^a W_{-1}^b v^{(1)\pm} -  9 f^{ab}_c W_1^c L_{-1} v^{(1)\pm} - 
f^{ab}_c L_0^n W_0^c v^{(1)\pm} = 0.$$
With (\ref{ldeterminesall}) and (\ref{wltollw}) this implies
\begin{eqnarray}
0 & = & L_0^n W_0^c v^{(1)\pm} + L_1 L_{-1} W_0^c v^{(1)\pm}\nonumber\\
  & \Longrightarrow & L_0^n v^{(1)\pm} = -\tilde{v}^{(0)\pm}.\label{lnilfix}
\end{eqnarray}
In order to fix the structure of $M$ completely, we now only have to
determine the action of $W^a_0$ on $v^{(1)\pm}$. With
(\ref{wcomm}) and (\ref{lwcommute}) one computes
\begin{eqnarray}
[W_0^a W_0^b] v^{(1)\pm} & = & 2 f^{ab}_c \left(-\frac{1}{5}W_0^c + 
\frac{12}{5}W^c_1 L_{-1} + \frac{6}{5}L_0W_0^c\right) v^{(1)\pm}\nonumber\\
& \stackrel{\mbox{(\ref{wltollw},\ref{lnilfix})}}{=} & 
    2 f^{ab}_c W_0^c v^{(1)\pm}\nonumber\\
& \Longrightarrow & \forall v \in M_1: [W_0^a W_0^b] v = 2 f^{ab}_c W_0^c v.
\end{eqnarray}
This means, that $v^{(1)\pm}$ can be \emph{chosen} (one always has the freedom
of choice $v^{(1)\pm} \mapsto v^{(1)\pm} + v, v\in M^{(0)}_1$) in such a way 
that $$W_0^a v^{(1)\pm} = 2\tau^a v^{(1)\pm}.$$
\par\noindent
This module, which was also found by M.R.~Gaberdiel and H.G.~Kausch in the
fusion of lowest weight \WWW-modules \cite{kauschrat}, will be
denoted by $V^{\WW*}_1$. Note, that admissibility fixes the
structure of the module completely and (in contrast to the pure Virasoro case)
the moduli space of generalized staggered modules therefore consists of one 
point only.
\par~\par\noindent
We now examine the possibilities to extend a GLWM with lowest weight $0$ by 
another GLWM with lowest weight $1$ yielding an indecomposable 
module of nilpotency length 1.
With equations (\ref{extendmeanszero}) and (\ref{ldeterminesall})
we conclude, that the general module of this form is given by
\[V^{\WW*}_1\, /\,
\UW.\left(\lambda v^{(0)+} + \mu v^{(0)-}\right),\;
(\lambda,\mu)\in\CC^2\backslash\{(0,0)\}.\]
Two of these modules are isomorphic if and only if 
$\lambda_1\mu_2 = \lambda_2\mu_1$.
Thus the moduli space of inequivalent modules of this form is
given by the Riemannian sphere $\CC\cup\{\infty\}$.

\subsection{Other modules}
Of course it is always possible to construct new modules from old ones
by taking direct sums and modding out submodules (as an example, the module
$(V^{\WW*}_1)^{(0)}$ was constructed in such a way from two copies of 
$\tilde{V}^\WW_0$). This naturally produces an infinite variety of 
\WWW-modules. The remaining question is, whether there are
\WWW-modules that are neither exhibited in the preceding subsections nor
can be constructed from such using the aforementioned operations.

\begin{define} Let $\mathcal{M}$ be the moduli space of all inequivalent 
\WWW-modules
together with the three operations `taking submodules' (Sub), 
`taking direct sums' (Sum)
and `modding out submodules' (Mod).
\end{define}

\begin{theorem}\label{thatsall}
The moduli space $\mathcal{M}$ is generated from the set
\[\left\{V^\WW_{-1/8}, V^{\WW*}_0, V^\WW_{3/8}, V^{\WW*}_1\right\}
\label{moduleset}\]
by the three operations (Sub), (Sum) and (Mod).
\end{theorem}

\begin{proof}
Let $M$ be a \WWW-module. Without loss of generality we can
assume $M$ to be indecomposable. 
First assume that one of the irreducible modules $M^k$ from the filtration 
(\ref{simplechain}) is either isomorphic to $V^\WW_{-1/8}$ or to 
$V^\WW_{3/8}$. As the levels of an indecomposable module are integer spaced, 
all of the $M^k$ are isomorphic to $V^\WW_{-1/8}$ or $V^\WW_{3/8}$,
respectively. Since there are no JLWMs with lowest weight $-1/8$ and
$3/8$, $M$ is isomorphic to $V^\WW_{-1/8}$ or $V^\WW_{3/8}$ and we are
through.
\par\noindent
Now assume, that $M$ is an admissible indecomposable \WWW-module, which is
not generated from the above set of modules (\ref{moduleset}). We will 
prove this to be impossible in three steps:
\begin{itemize}
\item[(a)]
We first investigate the case $\nlength(M) = 1$. If $M$ is not obtained by one
of the above operations, it must be an extension of some combination of lowest
weight modules with lowest weight 0 by one ore more lowest weight module(s)
with lowest weight 1. Without loss of generality we can assume $M$ to be
generated by the representatives $v^+, v^-$ of the lowest weight vectors of 
\emph{one} GLWM with lowest weight 1. 
Then equations (\ref{lwcommute}) and (\ref{lnilfix})
imply that $\UW L_{-1} v^\pm + \UW W^a_{-1} v^\pm$ is either isomorphic to
$V^\WW_0$ or isomorphic to $V^\WW_0 \oplus V^\WW_0$. $M$ can therefore be
constructed from one or more copies of $V^{\WW*}_1 / V^\WW_1$.

\item[(b)]
We now turn to the case $\nlength(M) = 2$. Without loss of generality we
may assume that the upper module of $M$ is indecomposable.
If $M^{(1)}$ is either a GLWM with lowest weight 0 or one with lowest
weight 1, we are through by subsections \ref{wjordan}, \ref{wstaggered}
and \ref{wgenstagg}. 
Thus assume $M^{(0)}$ to be an extension of 
some combination of GLWMs with lowest weight 0 by $V^\WW_1$. We will now
show that this is impossible.
Let $w^\pm$ be representatives of the lowest weight vectors of $V^\WW_1$.
We first remark, that $L_0^n w^\pm \not=0$, because otherwise
$L_0^n W^a_{-1} w^\pm = W^a_{-1} L_0^n w^\pm = 0$, 
$L_0^n L_{-1} w^\pm = L_{-1} L_0^n w^\pm = 0$ and thus $\nlength(M) = 1$.
If $L_0^n W^a_{-1} w^\pm = L_0^n L_{-1} w^\pm = 0$, $M$ contains a
JLWM with lowest weight 1, which is not admissible.
We conclude, that $M$ contains a JLWM with lowest weight 0 as a
submodule. Let $v$ be a representative of the lowest weight vector of the
upper weight 0 - GLWM. The extension by $V^\WW_1$ forces 
\[L_1 v = W_1^a v = 0\label{mustvanish}\] (c.f. \refeq{extendmeanszero}). 
On the other hand, $v$ is upper LWV of the above-mentioned JLWM,
which forces $\UW.v / \UW.L_0 v = \tilde{V}^\WW_0$ (note the difference to
$(\UW.v)^{(1)} = V^\WW_0$) in contradiction to (\ref{mustvanish}).

\item[(c)]
Last but not least we have to deal with the case $\nlength(M) > 2$.
We will show, that such a module cannot exist. To prove this it
suffices to examine the case $\nlength(M) = 3$, because any module
with higher nilpotency length would have to contain a submodule
with nilpotency length 3. By subsection \ref{wjordan},
such a module cannot be a JLWM. First note, that the submodule
$L_0^n.M \subset M$ must contain one or more
linearly independent vectors $v_i, i\in I,$
with $L_0^d v_i = 0$, $L_0^n v_i \not= 0$, since otherwise the
module $M / M^{(0)}$ cannot be admissible. Hence, the \WWW-module
$\tilde{M} := M / \UW.\langle L_0.v_i, i\in I \rangle$ must have
nilpotency length 2, because $L_0^n.M / \UW.\langle L_0^n.v_i, i\in I \rangle$
has nilpotency length 1 and 
$\UW.\langle v_i, i\in I \rangle \subset M^{(0)}$.
On the other hand, in $\tilde{M}$ we have 
$\forall i \in I: \UW v_i \equiv \tilde{V}^\WW_0$, which is impossible
for the lower module of an indecomposable nilpotency 2 module.
\end{itemize}
\qed
\end{proof}

Motivated by this result we slightly extend the usual definition of
`rational model' as follows:

\begin{define} \label{rationaldef}
A \textbf{rational model} of a \WW-algebra or \textbf{rational
\WW-algebra} is a \WW-algebra $\AAA$ which fulfills the following
condition:
There exists a finite set of $\AAA$-modules, from which all
other \AAA-modules can be obtained by taking submodules,
factor modules and direct sums.
\end{define}

\subsection{Characters and modular properties}
The character of a \WWW-module is defined to be its character as an
\LL-module.
The characters of the irreducible \WWW-modules have been known for quite
a while \cite{flohr1, flohr2, modconstruct} and are given by
\begin{myeqnarraylabel}{wirrchars}
\chi^{}_{V^\WW_0}(q) & = & 
\sum_{k=0}^\infty (2k+1) \chi^{}_{M(h_{4k+3,1},-2)}(q)
= \frac{1}{2 \eta(q)}(\Theta_{1,2}(q) + (\partial\Theta)_{1,2}(q)),\\
\chi^{}_{V^\WW_{-1/8}}(q) & = &
\sum_{k=0}^\infty (2k+1) \chi^{}_{M(h_{4k+2,1},-2)}(q)
= \frac{1}{\eta(q)}\Theta_{0,2}(q),\\
\chi^{}_{V^\WW_{3/8}}(q) & = &
\sum_{k=0}^\infty (2k+2) \chi^{}_{M(h_{4k+4,1},-2)}(q)
= \frac{1}{\eta(q)}\Theta_{2,2}(q),\\
\chi^{}_{V^\WW_1}(q) & = & 
\sum_{k=0}^\infty (2k+2) \chi^{}_{M(h_{4k+5,1},-2)}(q)
= \frac{1}{2 \eta(q)}(\Theta_{1,2}(q) - (\partial\Theta)_{1,2}(q)),
\end{myeqnarraylabel}
\noindent
where the $\Theta_{\lambda,k}$ are the Jacobi-Riemannian theta functions
$$\Theta_{\lambda,k}(q) = \sum_{n\in\ZZ} q^{\frac{(2kn+\lambda)^2}{4k}},$$
the $(\partial\Theta)_{\lambda,k}$ are the affine theta functions
$$(\partial\Theta)_{\lambda,k}(q) = 
\sum_{n\in\ZZ} (2kn+\lambda) q^{\frac{(2kn+\lambda)^2}{4k}}$$
and the $\chi^{}_{M(h,c)}$ are the characters of the irreducible
\LL-modules with lowest weight $h$ and central charge $c$
(c.f. corollary \ref{lirrchar}).
The theta functions, the affine theta functions and the eta function
transform under the action of the modular group $SL(2,\ZZ)$ represented
by $T: \tau \mapsto \tau+1$ and $S: \tau \mapsto -1/\tau$ as
\begin{myeqnarraylabel}{thetatransform}
(S \eta)(\tau) & = & \sqrt{-i\tau} \eta(\tau)\\
(T \eta)(\tau) & = & e^{\frac{i\pi}{12}} \eta(\tau)\\
(S \Theta_{\lambda,k})(\tau) & = & \sqrt{\frac{-i\tau}{2k}} 
\sum_{\lambda'=0}^{2k-1}
e^{i\pi\frac{\lambda\lambda'}{k}} \Theta_{\lambda',k}(\tau)\\
(T \Theta_{\lambda,k})(\tau) & = & e^{i\pi\frac{\lambda^2}{2k}} 
\Theta_{\lambda,k}(\tau)\\
(S (\partial\Theta)_{\lambda,k})(\tau) & = & \sqrt{\frac{1}{2k}} 
(-i\tau)^{\frac{3}{2}}
\sum_{\lambda'=1}^{2k-1} e^{i\pi\frac{\lambda\lambda'}{k}} 
(\partial\Theta)_{\lambda',k}(\tau)\\
(T (\partial\Theta)_{\lambda,k})(\tau) & = & e^{i\pi\frac{\lambda^2}{2k}}
(\partial\Theta)_{\lambda,k}(\tau).
\end{myeqnarraylabel}
\noindent
Using the results of the preceding subsections we readily
compute the characters
\begin{myeqnarraylabel}{wotherchars}
\chi^{}_{\tilde{V}^\WW_0}(q) & = & 
\chi^{}_{V^\WW_0}(q) + 2 \chi^{}_{V^\WW_1}(q)
= \frac{1}{2 \eta(q)}(3\Theta_{1,2}(q) - (\partial\Theta)_{1,2}(q)),\\
\chi^{}_{V^{\WW*}_0}(q) & = & 
\chi^{}_{\tilde{V}^\WW_0}(q) + \chi^{}_{V^\WW_0}(q)
= \frac{2}{\eta(q)}\Theta_{1,2}(q),\\
\chi^{}_{V^{\WW*}_1}(q) & = & 
\left(2 \chi^{}_{V^\WW_0}(q) + \chi^{}_{V^\WW_1}(q)\right) + 
\chi^{}_{V^\WW_1}(q)
= \frac{2}{\eta(q)}\Theta_{1,2}(q) = \chi^{}_{\tilde{V}^{\WW*}_0}(q).
\end{myeqnarraylabel}
\noindent
The characters $\chi^{}_{V^\WW_0}$ and $\chi^{}_{V^\WW_1}$
generate ``character'' functions with logarithmic terms in $q$ under the action
of the modular group, which we cannot interpret as characters of
\WWW-modules (in \cite{flohr2} it was attempted to interpret these
functions in terms of generalized characters). Anyway, with respect to
theorem \ref{thatsall} it seem to be more natural to view the \WWW-modules
\[\left\{V^{\WW*}_0, V^{\WW*}_1, V^\WW_{-1/8}, 
V^\WW_{3/8}\right\}\label{modulebase}\]
as the building blocks of our theory. In fact, the characters of these
modules display a much more agreeable behaviour under the action of
the modular group. With (\ref{thetatransform}) one calculates:
\begin{myeqnarraylabel}{goodchars}
S \chi^{}_{V^{\WW*}_0} & = & \chi^{}_{V^\WW_{-1/8}} - \chi^{}_{V^\WW_{3/8}}\\
S \chi^{}_{V^{\WW*}_1} & = & \chi^{}_{V^\WW_{-1/8}} - \chi^{}_{V^\WW_{3/8}}\\
S \chi^{}_{V^\WW_{-1/8}} & = & \frac{1}{2}\chi^{}_{V^\WW_{-1/8}} + 
    \frac{1}{2}\chi^{}_{V^\WW_{3/8}} + \frac{1}{2}\chi^{}_{V^{\WW*}_0}\\
S \chi^{}_{V^\WW_{3/8}} & = & \frac{1}{2}\chi^{}_{V^\WW_{-1/8}} + 
    \frac{1}{2}\chi^{}_{V^\WW_{3/8}} - \frac{1}{2}\chi^{}_{V^{\WW*}_0}\\
T \chi^{}_{V^{\WW*}_0} & = & e^{\frac{i \pi}{6}} \chi^{}_{V^{\WW*}_0}\\
T \chi^{}_{V^{\WW*}_1} & = & e^{\frac{i \pi}{6}} \chi^{}_{V^{\WW*}_1}\\
T \chi^{}_{V^\WW_{-1/8}} & = & e^{-\frac{i \pi}{12}} \chi^{}_{V^\WW_{-1/8}}\\
T \chi^{}_{V^\WW_{3/8}} & = & -e^{-\frac{i \pi}{12}} \chi^{}_{V^\WW_{3/8}}.
\end{myeqnarraylabel}
\noindent
Because of $\chi^{}_{V^{\WW*}_0} = \chi^{}_{V^{\WW*}_1}$, the definition of
the modular matrices is somewhat arbitrary. With the base (\ref{modulebase}) 
the most general ansatz is given by
\[S =
\left(\begin{array}{cccc}
\alpha & \beta & \frac{\gamma}{2} & -\frac{\delta}{2}\\
-\alpha & -\beta & \frac{1-\gamma}{2} & \frac{\delta-1}{2}\\
1 & 1 & \frac{1}{2} & \frac{1}{2}\\
-1 & -1 & \frac{1}{2} & \frac{1}{2}
\end{array}\right),\;
T =
\left(\begin{array}{cccc}
\mu e^{\frac{i \pi}{6}} & (1-\nu) e^{\frac{i \pi}{6}} & \lambda & \kappa\\
(1-\mu) e^{\frac{i \pi}{6}} & \nu e^{\frac{i \pi}{6}} & -\lambda & -\kappa\\
0 & 0 & e^{-\frac{i \pi}{12}} & 0\\
0 & 0 & 0 & -e^{-\frac{i \pi}{12}}
\end{array}\right).\]
\noindent
There are six solutions satisfying $S^4 = \OOne$, $(ST)^3 = S^2$ and 
the charge conjugation matrix $C = S^2$
being a permutation matrix:

$$S_1 = \left(\begin{array}{cccc}
\frac{i}{2} & -\frac{i}{2} & \frac{1}{4} & -\frac{1}{4}\\
-\frac{i}{2} & \frac{i}{2} & \frac{1}{4} & -\frac{1}{4}\\
1 & 1 & \frac{1}{2} & \frac{1}{2}\\
-1 & -1 & \frac{1}{2} & \frac{1}{2}
\end{array}\right),\;
T_1 = \left(\begin{array}{cccc}
0 & e^{i\pi/6} & 0 & 0\\
e^{i\pi/6} & 0 & 0 & 0\\
0 & 0 & e^{-i\pi/12} & 0\\
0 & 0 & 0 & -e^{-i\pi/12}
\end{array}\right)$$

$$S_2 = \left(\begin{array}{cccc}
\frac{i}{2} & -\frac{i}{2} & \frac{1}{4} & -\frac{1}{4}\\
-\frac{i}{2} & \frac{i}{2} & \frac{1}{4} & -\frac{1}{4}\\
1 & 1 & \frac{1}{2} & \frac{1}{2}\\
-1 & -1 & \frac{1}{2} & \frac{1}{2}
\end{array}\right),\;
T_2 = \left(\begin{array}{cccc}
\frac{3 + i\sqrt{3}}{4} e^{i\pi/6} & \frac{1 - i\sqrt{3}}{4} 
e^{i\pi/6} & 0 & 0\\
\frac{1 - i\sqrt{3}}{4} e^{i\pi/6} & \frac{3 + i\sqrt{3}}{4} 
e^{i\pi/6} & 0 & 0\\
0 & 0 & e^{-i\pi/12} & 0\\
0 & 0 & 0 & -e^{-i\pi/12}
\end{array}\right)$$

$$S_3 = \left(\begin{array}{cccc}
\frac{i}{2} & -\frac{i}{2} & \frac{1}{4} & -\frac{1}{4}\\
-\frac{i}{2} & \frac{i}{2} & \frac{1}{4} & -\frac{1}{4}\\
1 & 1 & \frac{1}{2} & \frac{1}{2}\\
-1 & -1 & \frac{1}{2} & \frac{1}{2}
\end{array}\right),\;
T_3 = \left(\begin{array}{cccc}
\frac{3 - i\sqrt{3}}{4} e^{i\pi/6} & \frac{1 + i\sqrt{3}}{4} 
e^{i\pi/6} & 0 & 0\\
\frac{1 + i\sqrt{3}}{4} e^{i\pi/6} & \frac{3 - i\sqrt{3}}{4} 
e^{i\pi/6} & 0 & 0\\
0 & 0 & e^{-i\pi/12} & 0\\
0 & 0 & 0 & -e^{-i\pi/12}
\end{array}\right)$$

$$S_4 = \left(\begin{array}{cccc}
-\frac{i}{2} & \frac{i}{2} & \frac{1}{4} & -\frac{1}{4}\\
\frac{i}{2} & -\frac{i}{2} & \frac{1}{4} & -\frac{1}{4}\\
1 & 1 & \frac{1}{2} & \frac{1}{2}\\
-1 & -1 & \frac{1}{2} & \frac{1}{2}
\end{array}\right),\;
T_4 = \left(\begin{array}{cccc}
e^{i\pi/6} & 0 & 0 & 0\\
0 & e^{i\pi/6} & 0 & 0\\
0 & 0 & e^{-i\pi/12} & 0\\
0 & 0 & 0 & -e^{-i\pi/12}
\end{array}\right)$$

$$S_5 = \left(\begin{array}{cccc}
-\frac{i}{2} & \frac{i}{2} & \frac{1}{4} & -\frac{1}{4}\\
\frac{i}{2} & -\frac{i}{2} & \frac{1}{4} & -\frac{1}{4}\\
1 & 1 & \frac{1}{2} & \frac{1}{2}\\
-1 & -1 & \frac{1}{2} & \frac{1}{2}
\end{array}\right),\;
T_5 = \left(\begin{array}{cccc}
\frac{1 + i\sqrt{3}}{4} e^{i\pi/6} & \frac{3 - i\sqrt{3}}{4} 
e^{i\pi/6} & 0 & 0\\
\frac{3 - i\sqrt{3}}{4} e^{i\pi/6} & \frac{1 + i\sqrt{3}}{4} 
e^{i\pi/6} & 0 & 0\\
0 & 0 & e^{-i\pi/12} & 0\\
0 & 0 & 0 & -e^{-i\pi/12}
\end{array}\right)$$

$$S_6 = \left(\begin{array}{cccc}
-\frac{i}{2} & \frac{i}{2} & \frac{1}{4} & -\frac{1}{4}\\
\frac{i}{2} & -\frac{i}{2} & \frac{1}{4} & -\frac{1}{4}\\
1 & 1 & \frac{1}{2} & \frac{1}{2}\\
-1 & -1 & \frac{1}{2} & \frac{1}{2}
\end{array}\right),\;
T_6 = \left(\begin{array}{cccc}
\frac{1 - i\sqrt{3}}{4} e^{i\pi/6} & \frac{3 + i\sqrt{3}}{4} 
e^{i\pi/6} & 0 & 0\\
\frac{3 + i\sqrt{3}}{4} e^{i\pi/6} & \frac{1 - i\sqrt{3}}{4} 
e^{i\pi/6} & 0 & 0\\
0 & 0 & e^{-i\pi/12} & 0\\
0 & 0 & 0 & -e^{-i\pi/12}
\end{array}\right)$$

$$C = \left(\begin{array}{cccc}
0 & 1 & 0 & 0\\
1 & 0 & 0 & 0\\
0 & 0 & 1 & 0\\
0 & 0 & 0 & 1
\end{array}\right).$$

\noindent
Note, that cases (1)--(3) only differ in the $T$-matrix and
cases (4)--(6) are just cases (1)--(3) with the roles of
$V^{\WW*}_0$ and $V^{\WW*}_1$ interchanged. 
\par\noindent
If all characters were linearly independent, 
invariance of the partition function
$$Z = \sum_{i,j} m_{ij} \bar{\chi^{}_i} \chi^{}_j, m_{ij} \in \NN^0,$$
\noindent
would force 
$S^t$ to be unitary with respect to the scalar product given by the matrix
$M := (m_{ij})_{i,j}$, i.e. $\bar{S} M S^t = M$. 
Since (in the base (\ref{modulebase})) $\chi^{}_1$ and $\chi^{}_2$ are the
same, we have a weaker restriction:
\begin{myeqnarray}
\sum_{i,j = 1,2}(\bar{S} M S^t)_{i,j} & = & \sum_{i,j = 1,2}(M)_{i,j}\\
\sum_{i = 1,2}(\bar{S} M S^t)_{i,j} & = & \sum_{i = 1,2}(M)_{i,j},\;j = 3,4\\
\sum_{j = 1,2}(\bar{S} M S^t)_{i,j} & = & \sum_{j = 1,2}(M)_{i,j},\;i = 3,4\\
(\bar{S} M S^t)_{i,j} & = & (M)_{i,j},\;i,j = 3,4.\\
\end{myeqnarray}
\noindent
In all six cases this, together with $m_{ij} \in \NN^0$, 
fixes $M$ to be of the form
$$M = \left(\begin{array}{cccc}
\alpha & \beta & 0 & 0\\
\beta & \gamma & 0 & 0\\
0 & 0 & 2 (\alpha + 2\beta + \gamma) & 0\\
0 & 0 & 0 & 2 (\alpha + 2\beta + \gamma)
\end{array}\right),\; \alpha,\beta,\gamma \in \NN^0,$$
\noindent
yielding the (up to a scalar factor) unique partition function
$$Z = 2 (\alpha + 2\beta + \gamma) 
\left(2 |\Theta_{1,2}|^2 + |\Theta_{0,2}|^2 + |\Theta_{2,2}|^2\right),$$
which is the partition function of the Gaussian model with central charge
$c = 1$ compactified on a circle of radius 1 (for some speculations
on the relations between nonunitary and unitary CFTs with the same
partition functions c.f. \cite{flohr3}).

\subsection{Fusion rules}
The modules (\ref{modulebase}) do indeed close under fusion, as was 
shown by M.R.~Gaberdiel and H.G. Kausch \cite{kauschrat}.
In the base (\ref{modulebase}) the fusion matrices $(N_i)_{j,k}$ 
($V_i \star V_j = \bigoplus_{k} (N_i)_{k,j} V_k$) are given by
\begin{myeqnarray}
N_1 = N_2 & = & \left(\begin{array}{cccc}
2 & 2 & 0 & 0\\
2 & 2 & 0 & 0\\
0 & 0 & 2 & 2\\
0 & 0 & 2 & 2
\end{array}\right),\\
N_3 & = & \left(\begin{array}{cccc}
0 & 0 & 1 & 0\\
0 & 0 & 0 & 1\\
2 & 2 & 0 & 0\\
2 & 2 & 0 & 0
\end{array}\right),\\
N_4 & = & \left(\begin{array}{cccc}
0 & 0 & 0 & 1\\
0 & 0 & 1 & 0\\
2 & 2 & 0 & 0\\
2 & 2 & 0 & 0
\end{array}\right).
\end{myeqnarray}

\noindent
It was conjectured by E.~Verlinde \cite{verlinde} that for 
any rational model the $S$-matrix diagonalizes the fusion
matrices
$$S^{-1} N_i^t S = D_i$$
\noindent
where $D_i$ is a diagonal matrix and
$$(D_i)_{jj} = \frac{S_{ij}}{S_{0,j}}$$
\noindent
(`0' is the index of the vacuum module).
Unfortunately, in our model \emph{neither} of the two different 
possible $S$-matrices diagonalizes the fusion rules, but transforms
them into block-diagonal form.
This was to be expected, since the vacuum module (with trivial 
fusion rules) only occurs as a \emph{submodule} of one
of our basic modules (\ref{modulebase}).
\par~\par\noindent
The quantum dimensions
\[d(V) := \lim_{i\tau \nearrow 0} 
\frac{\chi^{}_{V}(\tau)}{\chi^{}_{V_0^\WW}(\tau)},\]
can be calculated via
\[d(V_3) = \lim_{q\nearrow 1} 
\frac{\Theta_{0,2}(q)}{\Theta_{1,2}(q)+(\partial\Theta)_{1,2}(q)}
= \lim_{q\searrow 0} \frac{2(\Theta_{0,2}(q)-2\Theta_{1,2}(q)+\Theta_{2,2}(q))}
{\frac{1}{i\pi}\log q (\partial\Theta)_{1,2}(q) + 
\Theta_{0,2}(q)-\Theta_{2,2}(q))}
= 2\]
and
\[\frac{d(V_i)}{d(V_3)} = \lim_{i\tau \nearrow 0}
\frac{\sum_{j=1}^4 S_{j,i} \chi^{}_{V_j} (-\frac{1}{\tau})}
{\sum_{k=1}^4 S_{k,3} \chi^{}_k (-\frac{1}{\tau})}
= \lim_{q \rightarrow 0}
\frac{\sum_{j=1}^4 S_{j,i} \chi^{}_{j} (q)}
{\sum_{k=1}^4 S_{k,3} \chi^{}_k (q)}
= \frac{S_{3,i}}{S_{3,3}}.\]
\noindent
They are given by
\[d(V_0^{\WW*}) = d(V_1^{\WW*}) = 4,\; d(V_{-1/8}^\WW) = d(V_{3/8}^\WW) = 2.\]
As expected, they indeed transform multiplicatively under fusion.

\section{Conclusions and outlook}
As we have seen, many of the properties of arbitrary representations
of the Virasoro algebra can be deduced from lowest weight
representations. In particular, there are no new critical exponents which
do not occur in lowest weight modules. For the (simple)
case of Jordan Verma modules, their maximal preserving submodules
were determined, yielding a formula for the characters of
minimal Jordan lowest weight representations.
\par\noindent
For general staggered modules, we found strong restrictions on
their submodules and proved the moduli spaces $\mathcal{V}_{h,h'}$
to be one-dimensional vector spaces if there is an embedding 
$V(h',c) \rightarrow V(h,c)$, and to be empty otherwise.
\par\noindent
It remains an open question, whether it is possible to embed
a Jordan Verma module into another as a proper submodule of
the maximal preserving submodule. Connected to that, the
classification of the maximal preserving submodule of
a staggered modules with given lowest weights $h^1,h^2$ 
also is an open problem (its maximal \emph{proper} submodule is, of course,
either a Jordan Verma modules ($\alpha = 0$) or itself a staggered
submodule with lowest weights $h^2$ or $h^3$, c.f. section \ref{modspaces}).
\par\noindent
It is another problem, to extend the results on the moduli space of
staggered modules to higher ranks (e.g. in the rank 3 case,
$L_0^n v^{(2)}$ is \emph{not} necessarily singular, and therefore
the moduli space $\mathcal{V}_{h^1,h^2,h^3}$ is not the direct sum
of $\mathcal{V}_{h^1,h^2}$ and $\mathcal{V}_{h^2,h^3}$).
In addition, the structure of maximal preserving submodules
of Jordan Verma modules can be more complicated at higher rank.
At $c=-2, h=\frac{3}{8}$, e.g., the maximal preserving submodule of
a rank 3 JVM is a staggered module with lowest weights $\frac{35}{8}$,
$\frac{35}{8}$ and $\frac{99}{8}$, which is neither a JLWM nor a
strictly staggered module.
\par~\par\noindent
The representation theory of $\WW$-algebras in the logarithmic
regime was exemplarily studied for the case of $\WWW$ at $c=-2$
yielding finitely many representations from which all others
can be constructed by taking submodules, factor modules and
direct sums. These basic representations close under fusion
and their characters span a finite dimensional representation
of the modular group.
\par\noindent
Various reasons suggest that similar results will hold for the
whole series of triplet algebras $\WW(2,(2p-1)^3)$ at $c=c_{1,p}$,
but this is yet to be proven.


\section*{Acknowledgements}
It is a pleasure to thank W.~Nahm for useful discussions and support.
I would also like to thank his research group, namely
M.~R\"osgen, R.~Schimmrigk, K.~Wendland, A.Wi{\ss}kirchen, for many
helpful discussion and the both pleasant and productive atmosphere.
I am also grateful to M.~Flohr and H.G.~Kausch for useful discussions 
and comments.


\begin{appendix}
\section{Notations}
The modes of a field $\Phi$ with conformal weight $h_\Phi$ are
defined by
$$\Phi(z) = \sum_{n=-\infty}^\infty \Phi_n z^{n-h_\Phi}.$$
For two fields $\Phi,\Psi$ with conformal weights $h_\Phi,h_\Psi$
their normal ordered product is defined as
$$N(\Phi,\Psi):=\sum_{n=-\infty}^\infty N(\Phi,\Psi)_n z^{n-(h_\Phi+h_\Psi)}$$
with
$$N(\Phi,\Psi)_n:=\eps_{\Phi\Psi} \sum_{k=-\infty}^{h_\Psi-1} \Phi_{n-k}\Psi_k
+ \sum_{k = h_\Psi}^\infty \Psi_k\Phi_{n-k},$$
where $\eps_{\Phi\Psi}=-1$ if both $\Phi$ and $\Psi$ are fermionic and $+1$ in
all other cases.
\par\noindent
The quasiprimary normal ordered product $\Nop(\Phi,\Psi)$ is the projection
of $N(\Phi,\Psi)$ onto the space of quasiprimary fields. 
If $\{\Phi_i, i\in I\}$ is a base of the space of quasiprimary fields, it is 
explicitly given by the formula
\begin{eqnarray*}
\mathcal{N}(\Phi_j, \partial^n\Phi_i) & = &
\sum_{r=0}^{n} (-)^r {n \choose r} {{2(h_i+h_j+n-1)} \choose {r}}^{-1}
    {{2h_i+n-1} \choose {r}} \partial^r N^{(h_i+n-r)}
    (\Phi_j, \partial^{n-r}\Phi_i)\\
& & - (-)^n \sum\limits_{\{k | h_{ijk} \ge 1\}} C_{ij}^k {{h_{ijk}+n-1} 
    \choose {n}} {{2(h_i+h_j+n-1)} \choose {n}}^{-1}\\
& & \hspace{2cm}\times {{2h_i+n-1} \choose {h_{ijk}+n}} {{\sigma_{ijk}-1} 
    \choose {h_{ijk}-1}}^{-1} 
    \frac{\partial^{h_{ijk}+n}\Phi_k}{(\sigma_{ijk}+n)(h_{ijk}-1)!}
\end{eqnarray*}
where $\sigma_{ijk} := h_i + h_j + h_k - 1$, $h_{ijk} = h_i+h_j-h_k$ and
the $C_{ij}^k$ are the structure constants of the chiral algebra.
\par\noindent
With the above notations the isomorphism between the fields and the
vacuum representation is given by
\begin{eqnarray*}
\rho\left(\Phi\right) & = & \Phi_{h_\Phi} |0\rangle, \\
\rho^{-1}\left(\Phi_{i_1,n_1}\ldots\Phi_{i_k,n_k} |0\rangle\right) & = &
N(N(\ldots N(\Phi_{i_{k}}^{(n_{k})},
\Phi_{i_{k-1}}^{(n_{k-1})}),\ldots),
\Phi_{i_{1}}^{(n_{1})})
\end{eqnarray*}
where $h_i$ is the conformal weight of $\Phi^{(i)}$,
$\Phi_i^{(n)} := \frac{\partial^{n-h_{i}}}{(n-h_i)!}\Phi_i$ 
and $\forall k: n_i \ge h_i$.
\par~\par\noindent
For more details see e.g. \cite{nahmwconstruct}.
\end{appendix}


\end{document}